\newcommand{\be}[0]{\begin{equation}}
\newcommand{\ee}[0]{\end{equation}}
\newcommand{\ba}[0]{\begin{eqnarray}}
\newcommand{\ea}[0]{\end{eqnarray}}

\newcommand\GeV{\,\mbox{GeV}}

\documentclass[report]{JHEP3}

\usepackage{epsfig,multicol}

\newcommand{\pom}{{I\!\!P}}
\newcommand{\reg}{{I\!\!R}}
\newcommand{\xpom}{x_\pom}

\newcommand\fverb{\setbox\pippobox=\hbox\bgroup\verb}
\newcommand\fverbdo{\egroup\medskip\noindent%
            \fbox{\unhbox\pippobox}\ }
\newcommand\fverbit{\egroup\item[\fbox{\unhbox\pippobox}]}
\newbox\pippobox

\title{A global analysis of diffractive events at HERA
}

\author{ S. Taheri Monfared $^{(a,b)}$, Ali N. Khorramian $^{(a,b)}$ and S. Atashbar Tehrani $^{(b)}$

\\
\small{\it
$^{(a)}$ Physics Department, Semnan University, Semnan, Iran \\
$^{(b)}$ School of Particles and Accelerators, Institute for
Research in Fundamental Sciences (IPM), P.O.Box 19395-5531, Tehran,
Iran \\
 }
\\

    E-mail:  \email{sara.taheri@ipm.ir}, \email{khorramiana@theory.ipm.ac.ir}, \email{Atashbar@ipm.ir}}
\received{}        
\revised{}
\accepted{}        

\abstract{ We extract diffractive parton distribution functions (DPDFs) and diffractive structure functions from  the most recent H1 and ZEUS diffractive DIS data obtained by various methods.
We consider Pomeron as an object with parton distribution function, evolving according to the next-to-leading order (NLO) DGLAP equations within the framework of the `Fixed Flavour Number Scheme' (FFNS). Having performed a global fit analysis, we achieve a very good description of all available measurements by introducing a new set of quark distribution form for the Pomeron.
 We predict longitudinal and charm proton diffractive structure function as well. Our results are compared with other analysis from the literature.}

\begin{document}

\section{Introduction}
 Structure functions are key ingredient for deriving parton distribution functions (PDFs) in nucleons, which is important for any process which involves colliding hadrons. These PDFs allow us to predict cross sections at particle colliders. For the success of the physics program having precise knowledge of PDFs is required \cite{DeRoeck:2011na,Albrow:2008pn,Arneodo:2005kd}.
Strategies to extract the most precise PDFs to be used at the LHC, and questions on how to use future LHC data to further constrain the PDFs are discussed in the PDF4LHC \cite{PDF4LHC} forum.

The H1 and ZEUS collaborations presented their results on inclusive and various exclusive reactions, which is being actively studied by theorists and give access to a broader understanding of proton structure. Although data-taking there has been stopped, new results continue to appear.

One of their most important experimental results, working at a center of mass energy of about $300$~GeV is observation of a significant fraction, around 15\%, of large rapidity gap events in deep inelastic scattering (DIS) \cite{Ahmed:1994nw,Ahmed:1995ns,Adloff:1997sc,Aktas:2007bv,Derrick:1993xh,Derrick:1995wv,Breitweg:1998gc,
Chekanov:2002bz,:2009qja,Adloff:1998yg,Adloff:2001be,Aktas:2006hy,Aktas:2006hx,:2010kz,
Chekanov:2005vv,Chekanov:2008cw,Chekanov:2004hy,Chekanov:2008fh,Collaboration:2011ij,Royon:2006xw}.
%
In this event, which is called diffraction, an exchanged photon of virtuality $Q^2$ dissociates through its interaction with the proton at squared four momentum transfer $t$ to produce a hadronic system $X$ with mass $M_X$.

There are four different theoretical approaches to analyze diffractive data, which are described and compared in Ref \cite{Royon:2006by}. These are: the Pomeron Structure Function (PSF) model formulated in the framework
of Regge phenomenology, the Bartels-Ellis-Kowalski-W�usthoff (BEKW) two gluon exchange dipole model  \cite{Bartels:1998zy,Bartels:1998ea}, Bialas-Peschanski (BP) model based on the BFKL Pomeron approach \cite{Mueller:1994jq} and the saturation model of Golec-Biernat and W�usthoff (GBW) \cite{GolecBiernat:1998js}.
These four frameworks are based on completely different theoretical concepts.
The best description of all available measurements can be achieved with either the PSF based model or the BEKW approach \cite{Royon:2006by}. In addition, the PSF approach works good also when fitted to the ZEUS-$M_X$ data set, which is not the case for other models \cite{Royon:2006by,Sapeta:2006wp}.
 In Regge language, diffraction and the rapidity gaps which persist at high energy are associated with Pomeron
exchange; the structure of the Pomeron could then be clarified \cite{Introduction:Regge}.

The study of DESY experimental data based on the different methods provides influential information that shall use to achieve the best precision possible in extracting diffractive parton distribution functions (DPDFs). The purpose of the present study which is based on the PSF approach is to
discuss extensively the treatment of data sets, test the compatibility of the data obtained with various experimental methods and  bring deeper insights into Pomeron functional form.
We perform a NLO QCD analysis of the most available H1 and ZEUS observables in $\overline{\rm MS}$ scheme in the framework of Regge phenomenology and extract DPDFs.


The outline of the paper is to give an introduction to the theoretical framework adopted for the diffractive events  in Section \ref{Theoretical framework of the diffractive events}.  Section \ref{Extraction of parton densities in the Pomeron} provides a new technique of parameterization and present light and heavy formulation of the structure function.
Section \ref{Data Analysis} describes methods of selecting diffraction at HERA and tackles the technical issue of compatibility between different data sets.
We discuss our global fitting procedure and the method of minimization has been applied along with in Section \ref{QCD fit}.  Our results are presented in Section \ref{Discussion of fit results}.
 Section \ref{Conclusion } contains conclusion and introduces a \texttt{FORTRAN}-code which is available.

\section{Theoretical framework of the diffractive events }
\label{Theoretical framework of the diffractive events}
\subsection{Diffractive cross sections}
 The data are often presented in the form of a $t$-integrated reduced diffractive
neutral current cross section
$\sigma_{r}^{D(3)}$, defined via
\begin{equation}
\frac{d^3\sigma^{ep \rightarrow eXY}}{{\rm d} x_\pom \ {\rm d} \beta \ {\rm d} Q^2} =
 \frac{4\pi \alpha^2}{x Q^4}\left ( 1 - y + \frac{y^2}{2} \right )
\sigma_{r}^{D(3)}(\xpom,\beta,Q^2) \ ,
\label{sigmar}
\end{equation}
or in terms of a diffractive
structure function $F_2^{D(3)}(x_\pom,\beta,Q^2)$ \cite{Newman:2005wm}
\begin{equation}
\sigma_r^{D(3)} = F_2^{D(3)} - \frac{y^2}{Y_+} F_L^{D(3)} \ ,
\label{eq:sigf2fl}
\end{equation}
where  $Y_+=1+(1-y)^2$.
Due to the presence of the $ {y^2}/{Y_+}$ factor, the second term in (\ref{eq:sigf2fl}) can be neglected anywhere but at very large $y$ \cite{Favart:2009zz,Lohr:2008zz}. Since the kinematical variables are bound by $sxy=Q^2$,
the longitudinal structure function plays an important role in low-$x$ physics.
 The effects of $F_2^{D(3)}$ and $F_L^{D(3)}$ are considered through their NLO dependence on DPDFs which will be shown in next section.

\subsection{QCD hard scattering factorization}

The proof that QCD hard scattering factorization can be applied to diffractive DIS \cite{collins} indicates that the cross section for the diffractive process can be considered in terms of convolutions of universal partonic cross sections $\hat{\sigma}^{e i} (x, Q^2)$, which are the same as those for DIS,  with DPDFs $f_i^D$ as
\begin{equation}
{\rm d} \sigma^{ep \rightarrow eXY} (x, Q^2, \xpom, t) =
 \sum_i \ f_i^D(x, Q^2, \xpom, t) \ \otimes \
{\rm d} \hat{\sigma}^{ei}(x,Q^2) \ .
\label{equ:diffpdf}
\end{equation}
Moreover, it is assumed that the shape of the DPDFs is not dependent to $\xpom$ and $t$ and their normalisation is governed by Regge asymptotics \cite{Aktas:2006hy,ingschl}.
The mentioned assumption is required and compatible with the data fitted.
The DPDFs can then be presented as a sum of two factorised contributions corresponding to Pomeron and sub-leading Reggeon:
\begin{equation}
f_i^D(x,Q^2,\xpom,t) = f_{\pom/p}(\xpom,t) \cdot f_i (\beta,Q^2)+
n_\reg \cdot f_{\reg/p}(\xpom,t) \cdot f_i^\reg (\beta,Q^2) \ .
\label{reggefac2}
\end{equation}
$f_i (\beta,Q^2)$ and $f_i^\reg (\beta,Q^2)$ are the partonic structure of Pomeron and Reggeon, respectively.
Reggeon contribution contributes significantly only at low $\beta$ and large $\xpom$.
To have detailed information on the flux factor refer to Refs. \cite{Aktas:2006hy,Aktas:2006hx,h1f2d94}.

\section{Extraction of parton densities in the Pomeron}
\label{Extraction of parton densities in the Pomeron}
In comparison to the inclusive parton distribution, the DPDFs contain two additional variables ($\xpom,t$). Since they do not influence evolution, we focus our attention on the term depending on $\beta$ and $Q^2$.

\subsection{Parametrization }
\label{Parametrization }
Due to the availability of new high precision measurements, phenomenological  groups try to provide for precise understanding of the nucleon structure and its partonic content.
Following our previous studies on extraction of precise polarized and unpolarized PDFs \cite{Khorramian:2010qa,Khorramian:2009xz,Khorramian:2008yh,Atashbar Tehrani:2007be,Khorramian:2006wg,Khorramian:2004ih}, we have motivation to tackle with the parameterization of DPDFs.

A simple instruction is adopted in which the parton distributions of  both the Pomeron and the Reggeon are parametrised in terms of non-perturbative input distributions as a function of $z$ at $Q_0^2$. \\
The structure of sub-leading trajectory
 $f_i^\reg$ are obtained from a parameterisation extracted from fits to pion
structure function data with a free normalization to be determined by the diffractive data.
We test the dependence on the pion structure function by
performing our fitting procedure base on the the GRV parametrisation \cite{grvpion} instead on the Owens one \cite{owens}.
No considerable deviation exists between both fits (similar result is reported in \cite{Royon:2006by}).

 We parametrize the Pomeron,  a light flavour singlet distribution $\Sigma(z)$, consisting of $u$, $d$ and $s$
quarks and anti-quarks with $u=d=s=\bar{u}=\bar{d}=\bar{s}$, and a gluon distribution $g(z)$ at an initial scale $Q_0^2$=3 GeV$^2$, such that
\begin{equation}
z \Sigma (z,Q_0^2) = A_{\Sigma} \, z^{B_{\Sigma}} \, (1 - z)^{C_{\Sigma}}(1+D_{\Sigma} z+E_{\Sigma} z^{F_{\Sigma}}) \ ,
\label{param:general1}
\end{equation}
\begin{equation}
z g(z,Q_0^2) = A_g \, z^{B_{g}} \,  (1 - z)^{C_g}~  e^{- \frac{a}{1-z}} \ ,
\label{param:general2}
\end{equation}
with $a=0.01$.
Here $z$ is the fractional momentum of Pomeron carried by the struck parton.
The DPDFs can be extracted from fits to the data applying the DGLAP \cite{dglap} evolution equations.
We present a new set of quark singlet distribution form which will be discussed briefly latter.

The gluon density is poorly constrained by the data, therefore it found to be insensitive to the $B_{g}$ parameter.
Performing a fit without the parameter $C_g$ confirms the lack of sensitivity of gluon distribution to large $z$.
Finally, we consider the parameterization of the gluon density identical to that used in the second analysis of \cite{Aktas:2006hy}.
Our motivation to apply this simple form of gluon distribution is that predictions based on it give the better description of diffractive dijet production cross sections measured in DIS in comparison with the form including $C_g$ parameter \cite{Aktas:2007bv}.


We studied the value of $a$ and found that it is not fundamental, e.g. $a=0.001$ also results in an acceptable fit. Moreover, we performed the analysis with $a=0$ considering the behavior at high $z$ by substituting $(1-z)^{C_g}$. We did not get any satisfactory results.
It means that although the exponential term is large in a region where no data are considered in the fit, the form of the parameterization for the gluon distribution at large $z$ is important \cite{Royon:2000wg}.

\subsection{Light flavour contribution}
\label{Light flavour contribution}
The Eqs.~(\ref{sigmar}) and (\ref{eq:sigf2fl}) are written in analogy with the way ${d^2\sigma^{ep \rightarrow eX}}/{{\rm d} x_{Bj} \ {\rm d} Q^2}$ is related to the structure functions $F_2$ and $F_L$ for inclusive DIS. Similarly, diffractive structure function can be described by

\begin{equation}
F_i(\beta,Q^2)=\sum_j C_i^j (\beta,\frac{Q^2}{\mu^2})\otimes f_j(\beta,\mu^2) \ ,
\label{param:general3}
\end{equation}
to all order in perturbation theory. Here $\mu^2$ denotes the factorization scale. As outlined earlier $f_j$ is considered in terms of the  light flavour singlet distribution $\Sigma(z)$ and  gluon distribution $g(z)$.
Generally speaking, structure function can be written  like
\begin{equation}
F_i(\beta,Q^2)=F_i^{light}(\beta,Q^2)+F_i^{heavy}(\beta,Q^2,m_h^2) \ .
\label{param:general4}
\end{equation}
The flavour singlet contribution up to NLO is given by \cite{Gluck:2006pm,Gluck:2007sq}
\begin{equation}
\frac{1}{x}F_i^{light}(\beta,Q^2)=\frac {2}{9} \biggl( C_{i,q} \otimes \Sigma + C_{i,g} \otimes g \biggr) (\beta,Q^2) ,
\label{param:general5}
\end{equation}
where $i = 2,L$ and the $C_{i,q}$ and $C_{i,g}$ are the common NLO coefficient functions \cite{ Gluck:2007sq,vanNeerven:2000uj}.
\subsection{Heavy flavour contribution}
\label{Heavy flavour contribution}

The treatment of heavy quarks is something that nearly every group does slightly and it can lead to surprisingly different results for PDFs extracted \cite{DeRoeck:2011na}.
We consider the effects of a heavy quark within the framework of the so called FFNS where, beside the gluon, only the light quarks $q={u,d,s}$ are considered as `intrinsic' genuine partons, i.e. massless partons within the nucleon, and heavy quarks $h={c,b,t}$ should not be included in the parton structure of the nucleon. They are always produced in the final state from the initial light quarks and gluons in near threshold region, i.e. $Q^2 \sim m_h^2$ \cite{Gluck:1980cp,Gluck:2008gs}. However, even for very large values of $Q^2$, $Q^2 \gg m_{h}^2$, these FFNS predictions are in remarkable agreement with DIS data \cite{Gluck:2007ck,Gluck:1998xa}.

The heavy structure functions are given through  \cite{Ajaltouni:2009zz}
\begin{equation}
F_i^h(\beta,Q^2)=\sum_k  C_{i,k}^{FF,n_f} (Q^2/ m_h^2)\otimes f_{i,k}^{n_f}(Q^2) \ .
\label{param:general6}
\end{equation}
Here, all quark flavours below $m_h$ are treated as zero-mass and one sums over $k=g,u,\bar u, d, \bar d, ...$ up to $n_f$ flavours of light (massless quarks). The mass of heavy quark, $m_h$, appears explicitly in the Wilson coefficients $C_{i,k}^{FF,n_f}$, as indicated in Eq.  (\ref{param:general6}).

 No simple analytic expressions can be given for the coefficient functions. To consider heavy contribution of $F_{2,L}$ we follow the standard manner applied in \cite{Riemersma:1994hv}

\begin{eqnarray}
F_{k}(\beta,Q^2,m^2) &=&
\frac{Q^2 \alpha_s}{4\pi^2 m^2}
\int_{\beta}^{z_{\rm max}} \frac{dz}{z}  \Big[ \,e_H^2 f_g(\frac{\beta}{z},\mu^2)
 c^{(0)}_{k,g} \,\Big] \nonumber \\&&
+\frac{Q^2 \alpha_s^2}{\pi m^2}
\int_{\beta}^{z_{\rm max}} \frac{dz}{z}  \Big[ \,e_H^2 f_g(\frac{\beta}{z},\mu^2)
 (c^{(1)}_{k,g} + \bar c^{(1)}_{k,g} \ln \frac{\mu^2}{m^2}) \nonumber \\ &&
+\sum_{i=q,\bar q} \Big[ e_H^2\,f_i(\frac{\beta}{z},\mu^2)
 (c^{(1)}_{k,i} + \bar c^{(1)}_{k,i} \ln \frac{\mu^2}{m^2}) \nonumber \\ &&
+ e^2_{L,i}\, f_i(\frac{\beta}{z},\mu^2) (d^{(1)}_{k,i} + \bar d^{(1)}_{k,i}
\ln\frac{\mu^2}{m^2}) \, \Big]  \,\Big] \,,
\end{eqnarray}
where $k = 2,L$ and the upper boundary on the integration is given by
$z_{\rm max} = Q^2/(Q^2+4m^2)$. Further $f_i(x,\mu^2), (i=g,q,\bar q)$
denote the parton densities in the proton. $e_H^2$ and $e_L^2$ represent the charge squared of the heavy and light quarks respectively.
 The coefficient functions, represented by $c^{(l)}_{k,i}(\eta, \xi)\,,\bar c^{(l)}_{k,i}
(\eta, \xi)\,$ and $d^{(l)}_{k,i}(\eta, \xi)\,,\bar d^{(l)}_{k,i}(\eta, \xi)$
are fully described in \cite{lrsn1,Harris:1995tu} up to NLO. Some progress at next-to-next-to-leading order (NNLO) can be found in \cite{Bierenbaum:2009mv}.

\section{Data Analysis}
\label{Data Analysis}

\subsection{Methods of selecting diffraction at HERA }
\label{Methods of selecting diffraction at HERA}
Cross sections in deep inelastic diffractive scattering are not uniquely defined.
 Different methods exist to select diffractive events. These methods select samples which contain different fractions of proton dissociative events.  Cross sections are not always given with corrections for proton dissociation.
Three distinct methods of
\begin{itemize}
\item{  Forward (Leading) Proton Spectrometer Method,}
\item{Large Rapidity Gap Method,}
\item {$M_X$ Method,}
\end{itemize}
have been employed  by the H1 and ZEUS experiments, which select inclusive diffractive events of type $ep \rightarrow eXY$. The advantages and disadvantages of each method are explained in detailed in \cite{Newman:2005wm,Lohr:2008zz}.

A reasonable compatibility between these techniques and between H1 and ZEUS results have been observed, which shows that there is no strong bias between these experimental techniques. As can be seen in several experimental references \cite{Royon:2006xw,Newman:2005wm,Schoeffel:2008fj,Newman:2009km, Ruspa:2008qj,Capua:2011sm}, there are tolerable agreement and common fundamental features between different data sets for much of the kinematic range. However, there are clear regions of disagreement especially between  the $M_X$ and LRG Methods  \cite{Abramowicz:2005yc,Watt:2005gt}. \\
The full HERA data sample analysis is a powerful technique to achieve the best  precision possible in extracting DPDFs.
First steps are taken towards the combination of the H1 and ZEUS results \cite{Newman:2009km}. However, already at the present level, much can be done with existing data for the understanding of diffraction at HERA.
Our strategy to study the HERA data will be discussed in Section \ref{Our strategy on data combination}.
%
\subsection{Data sets }
\label{Data sets}
In this Section, we describe the different available H1 and ZEUS data sets.
Since the various data sets correspond to different ranges in the outgoing proton system mass, $M_Y$,
i.e. $M_Y=m_p$ in case of FPS (LPS) results and $M_Y< 2.3$ GeV in case of the $M_X$ results,
additional factors are required before comparisons can be made.
For all data and fit comparisons, all data are transported to the H1-LRG-06 \cite{Aktas:2006hy} measurement range $M_Y < 1.6$ GeV.
The scaling factors, which has been obtained by corresponding experimental groups, are independent of $\beta$, $Q^2$ and $\xpom$ \cite{Newman:2009km,Ruspa:2008qj}.
As it was discussed in the experimental analysis, this is simply impossible to access any sensitivity of proton dissociation with respect to kinematics.
 To summarize, we give the H1 and ZEUS available different data sets together with their scaling factors in the following.

\begin{itemize}    
\item $\sigma_{r}^{D}$ measured by H1 collaboration using the Large Rapidity Gap Method labeled H1-LRG-06 \cite{Aktas:2006hy}, this is the default data set which is not corrected further. The analysis is restricted to the region $ y < 0.9$;

\item $\sigma_{r}^{D}$ measured by H1 collaboration with the leading final state proton detected in Forward Proton Spectrometer  labeled H1-FPS-06 \cite{Aktas:2006hx}, data multiplied by the global factor 1.23. The analysis is restricted to the region $ y < 0.5$;

\item $\sigma_{r}^{D}$ measured by H1 collaboration with the leading final state proton detected in Forward Proton Spectrometer  labeled H1-FPS-10 \cite{:2010kz}, data multiplied by the global factor 1.20. The analysis is restricted to the region $ y < 0.7$;

\item $F_2^{D}$ measured by ZEUS collaboration with the leading final state proton detected in Leading Proton Spectrometer  labeled ZEUS-LPS-04 \cite{Chekanov:2005vv}, data multiplied by the global factor 1/0.75=1.33;

\item $\sigma_{r}^{D}$ measured by ZEUS collaboration with the leading final state proton detected in Leading Proton Spectrometer  labeled ZEUS-LPS-09 \cite{Chekanov:2008cw}, data multiplied by the global factor  1/0.75=1.33. The analysis is restricted to the region $ y < 0.5$;

\item $\sigma_{r}^{D}$ measured by ZEUS collaboration using the Large Rapidity Gap Method labeled ZEUS-LRG-09 \cite{Chekanov:2008cw}, data multiplied by the global factor 1.05 \footnote{In Ref.~\cite{Chekanov:2008cw}, the ZEUS cross section is quoted and given in the table at the proton mass. This implies subtracting the proton dissociation background by applying a constant scaling factor 0.75. To correct them to the $M_y$ range of H1-LRG-06, $M_y < 1.6$ GeV, one has to multiply the cross section, before the subtraction of the proton dissociation background, by the factor  0.91. Moreover, the ZEUS data are higher than H1 by 13\% on average. Therefore the global factor 0.91$\times$(1-0.13)$\times$(1/0.75) $\simeq$1.05  must be applied. }.
The analysis is restricted to the region $ y < 0.6$;

\item $F_2^{D}$ measured by ZEUS collaboration using the $M_X$ Method labeled ZEUS-$M_X$-05 \cite{Chekanov:2004hy}, data multiplied by the global factor 0.88 \footnote{In Ref.~\cite{Chekanov:2008cw}, $M_X$ results are normalized to ZEUS-LRG-09 results by the factor of 0.83. Consequently, to transport $M_X$ data to the H1-LRG-06 measurement the factor of 0.83$\times$0.91$\times$(1-0.13)$\times$(1/0.75)$\simeq$0.88 must be applied.};

\item $F_2^{D}$ measured by ZEUS collaboration using the $M_X$ Method labeled ZEUS-$M_X$-08 \cite{Chekanov:2008fh}, data multiplied by the global factor 0.88;

\item $F_{2,L}^{D(3)}$ and $\sigma_{r}^{D(3)}$ recently measured by H1 collaboration using the LRG Method labeled H1-LRG-11 \cite{Collaboration:2011ij}, cross sections  data with proton beam energy of $E_p$=460, 575 and 920 $\GeV$ multiplied by the global factors of 0.97, 0.99, 0.97, respectively. The analysis is restricted to the region $ y < 0.85$;
\end{itemize}

\subsection{Our strategy on data combination}
\label{Our strategy on data combination}
There are several issues on data analysing that motivated us to try out a deeper look in comparison with what has been explained up to now.
\begin{itemize}
\item The first challenge is that $M_X$, used by ZEUS to separate diffractive from non-diffractive events, and LRG methods use the same data and thus they are strongly correlated \cite{:2009qja,Ruspa:2008qj}. 
\item Although the leading-proton data previously suffered from low statistics and hence were unlikely to have much influence on the fit results, the high statistics of the present data make them competitive in precision with the result of the LRG method \cite{:2010kz}.
\item The LRG results from H1 and ZEUS are compatible in most of the kinematic region covered by measurements \cite{Newman:2009km}.
\end{itemize}
Due to the above items, data based on the $M_X$ method has not been considered in our global analysis.
To study the influence and compatibility of the leading-proton data, we perform two scenarios: 1) based on the both LRG and FPS/LPS data and 2.) only LRG data.
Comparison between two scenarios shows consistency of leading-proton data and confirms the importance of including them to increase the fit precision.\\
The total data sets that we use in the present global analyses together with their type and $\beta, \xpom$ and $Q^2$ range are listed in Table~\ref{tab:data}.
The fitted normalization ${\cal N}_{i}$ of the data sets included in the global fit, for each data set $i$ are also shown in this table.
%
%
\vspace{2mm} \noindent \\
\begin{table*}[h]
\begin{center}
\begin{tabular}{|c|c|c|c|c|c|c|} 
\hline\hline Lable & Data set& $\beta$-range & $\xpom$-range & Q$^{2}$-range &
 N$_{Data}$ & \textbf{${\cal {N}}_{i}$}    
\\ \hline\hline
H1-LRG-06 & $\sigma_{r}^{D(3)}$ & 0.004-0.8 & 0.001-0.03  &  8.5-1600 & 190  \cite{Aktas:2006hy} &   0.9958     \\
H1-FPS-06 &$\sigma_{r}^{D(3)}$& 0.02-0.7 & 0.0011-0.08 & 10.7-24 & 40  \cite{Aktas:2006hx} &    1.0023   \\
H1-FPS-10 &$\sigma_{r}^{D(3)}$& 0.006-0.562 & 0.0025-0.075 & 8.8-200 & 100   \cite{:2010kz}  &  1.0002    \\
ZEUS-LPS-04 &$F_2^{D(3)}$& 0.007-0.48 & 0.0005-0.06 & 13.5-39 & 27  \cite{Chekanov:2005vv}  &    1.0005     \\
ZEUS-LPS-09 &$\sigma_{r}^{D(3)}$& 0.013-0.609 & 0.0009-0.09 & 14-40 & 42  \cite{Chekanov:2008cw}  &   0.9845   \\
ZEUS-LRG-09 &$\sigma_{r}^{D(3)}$& 0.025-0.795 & 0.0005-0.014 & 8.5-225 & 155  \cite{Chekanov:2008cw}  &    1.0094    \\
H1-LRG-11 &$F_{2,L}^{D(3)}$& 0.089-0.699 & 0.0005-0.003 & 11.5-44 & 20 \cite{Collaboration:2011ij}  &    0.9739    \\
H1-LRG-11 &$~\sigma_{r, (\sqrt{s}=225,252)}^{D(3)}$& 0.089-0.699 & 0.0005-0.003 & 11.5-44 & 25  \cite{Collaboration:2011ij}  &    0.9739  \\
H1-LRG-11 &$~\sigma_{r, (\sqrt{s}=319)}^{D(3)}$& 0.089-0.699 & 0.0005-0.003 & 11.5-44 & 12  \cite{Collaboration:2011ij}  &   0.9608   \\
H1-LRG-11 &$~R^{D}$& 0.089-0.699 & 0.0005-0.003 & 11.5-44 & 10 \cite{Collaboration:2011ij}  &   1   \\
\hline
Total &  &  & & & 621  &        \\ 
\hline\hline
\end{tabular}
\vspace{2mm} \noindent \\
\caption{{Overview of the published data points for $\beta \leqslant 0.8$, $M_x > 2 \GeV$ and $Q^2 \geqslant 8.5$ GeV$^2$  together with the fitted normalization shifts  \textbf{${\cal {N}}_{i}$}. The normalisation uncertainty cancels in the ratio $~R^{D}$.
 \label{tab:data}}}
\end{center}
\end{table*}
\section{QCD fit}
\label{QCD fit}
Before getting into the detailed procedure of our QCD fit, let us mention a few points.
 Following the treatment of  \cite{Aktas:2006hy}, in order to avoid regions which are most likely to be influenced by higher twist contributions or other problems with the chosen theoretical framework, only data in the range  $\beta \leqslant 0.8$, $M_x > 2 \GeV$ and $Q^2 \geqslant 8.5$ GeV$^2$ are considered in the fit. We found that although individual fitting procedure of some of the data sets works reasonably well at lower $Q^2$, the quality of the global fit with all available data sets drops for $Q^2 < 8.5$ GeV$^2$.
 The effect of $F_L^D$ are considered through its relation to the NLO parton densities, such that no explicit cut on $y$ is required.

The different fitting groups use various values of the charm and bottom quark masses, which may change PDFs up to a couple of percent  \cite{DeRoeck:2011na}. MSTW \cite{Martin:2010db} have recently presented a detailed study of mass dependence in the PDFs.
Here, the heavy quark masses are set to world average values of  $m_c=1.41\GeV$, $m_b=4.50\GeV$ and $m_t=175\GeV$ \cite{Nakamura:2010zzi}. The strong coupling is set via
$\alpha_s^{(3)}(Q_0^2)=0.306$ for three flavours, which corresponds  to the world average
$\alpha^{(5)}_s (M_Z^2)=0.1184$ for five flavours  \cite{mst}.
Our analysis is performed using the QCD-PEGASUS program \cite{Vogt:2004ns}.
We work at NLO in the QCD evolution and take
the renormalization and factorization scales to be equal $(\mu_{R}=\mu_{F})$.
An N-space evolution of Eq. (\ref{param:general2}) would require an accurate internal re-parametrization which facilitates a computation of the moments on the Mellin inversion contour.
\subsection{Detailed procedure}
\label{Detailed procedure}

We step in to the process of this project by performing a QCD fit under the same conditions and conventions as in fit B of Ref.  \cite{Aktas:2006hy}.  Having considered the same functional form
 and cut scenario for 190 data points, we obtain almost identical result in input scale of 2.5 GeV$^2$.
\\
In a second step, we combine all the data sets with their normalization factors as defined in Section \ref{Data sets}.
 It leads to  a $\chi ^{2}/$dof$=689.3/610=1.13$. 
\\
Finally,  we find that the exponential term plays negligible role in quark singlet distribution.
Moreover, considering the additional factor of $(1+D_{\Sigma} z+E_{\Sigma} z^{F_{\Sigma}})$ provides flexibility to obtain a good description of the data.
 Thus, we vary $\Sigma$ distribution to the functional form of Eq. (\ref{param:general1}) to improve our fitting procedure.
 This specific choice of parameterization reduces $\chi^2/$dof value  from $1.13$ to $0.99$.

As already discussed in \cite{Aktas:2006hy,Adloff:2000qk}, if the number of parameters in describing parton densities is small, the $\chi^2$ as well as the fit values would be sensitive to the choice of the input scale.
However, it is expected that the fit result to be independent of $Q_0^2$.
Thus, another advantage of using the new form of DPDF, introduced in this paper, is that the fit result is independent of varying the input scale in the reasonable range (In the order of 1 $\GeV^{2}$).

$\alpha_{\pom}(0)$ is considered as a free parameter in the fit.
Since the experimental treatments and cuts are different for each data set, we assign a global Reggeon normalization parameter to each data set for which this contribution is required, similar method is applied in Ref. \cite{Royon:2006by}.
%
Considering Reggeon normalization parameter to H1-LRG-11 is not needed because of its low $\xpom$ coverage.
 This gives us a total number of 14 unknown parameters which are presented in Table~\ref{tab:parameter}.

Our result for $\alpha_\pom (0)$ is compatible with the pomeron intercept describing soft hadronic scattering,
$\alpha_{\pom}(0) \simeq 1.08$ \cite{softpom,softpom2,softpom3}. It is also consistent with that
obtained from H1 data previously measured using the LRG and FPS
methods \cite{Aktas:2006hy,Aktas:2006hx,:2010kz} and with the ZEUS
measurements \cite{Chekanov:2004hy,Chekanov:2008fh}.

Considering the general form of  Eq. (\ref{param:general1}) for non-diffractive PDFs, in non-singlet combinations, e.g. the valence quarks and $(\bar{u}-\bar{d})$, $B$ parameter is expected to be $\sim$ 0.5. In singlet contributions, e.g. the sea and gluon, $B$ is expected to be $\sim$ 0 \cite{DeRoeck:2011na}. Here, in analogy with non-diffractive PDFs, we achieved similar amount for $B_{\Sigma}$.

\begin{figure}[t]
\centerline{\includegraphics[width=0.7\textwidth]{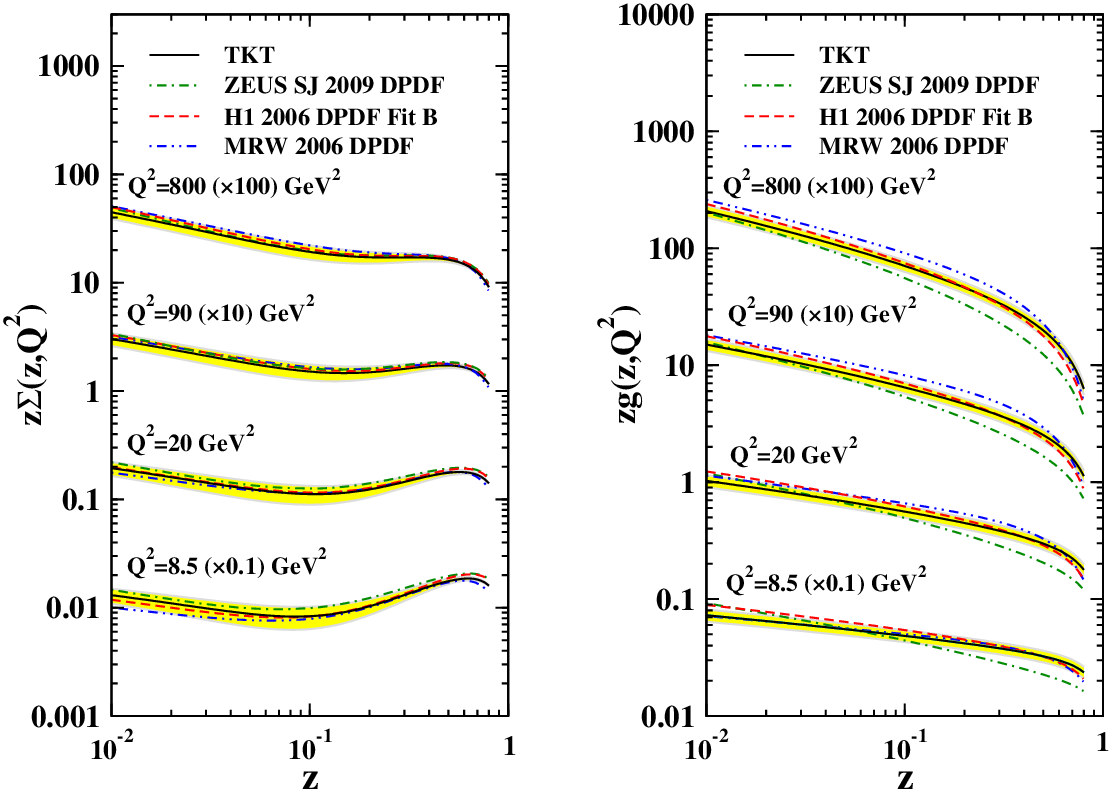}}
\caption{Comparison between the total quark singlet and gluon distributions obtained from our model (solid curve),  H1 2006 DPDF Fit B  (dashed curve) \cite{Aktas:2006hy}, ZEUS SJ (dashesd-dotted curve) \cite{:2009qja} and  MRW (dashesd-dotted-dotted curve) \cite{Martin:2006td}  at four different values of $Q^2$ as a function of $z$. The ZEUS SJ and MRW DPDFs plotted here are normalised to  $M_Y<1.6 \GeV$ by multiplying by a factor 1.23 relative to $M_Y=m_p$.}
\label{combined}
\end{figure}

The diffractive quark singlet and gluon distributions from our model are compared with the results from H1 2006 Fit B \cite{Aktas:2006hy}, MRW \cite{Martin:2006td} and ZEUS SJ \cite{:2009qja} for different values of $Q^2$ on a logarithmic $z$ scale in Figure \ref{combined}.  At low $Q^2$, both the quark singlet and the gluon densities remain large up to the highest $z$ values accessed.

\begin{figure}[b]
\centerline{\includegraphics[width=0.7\textwidth]{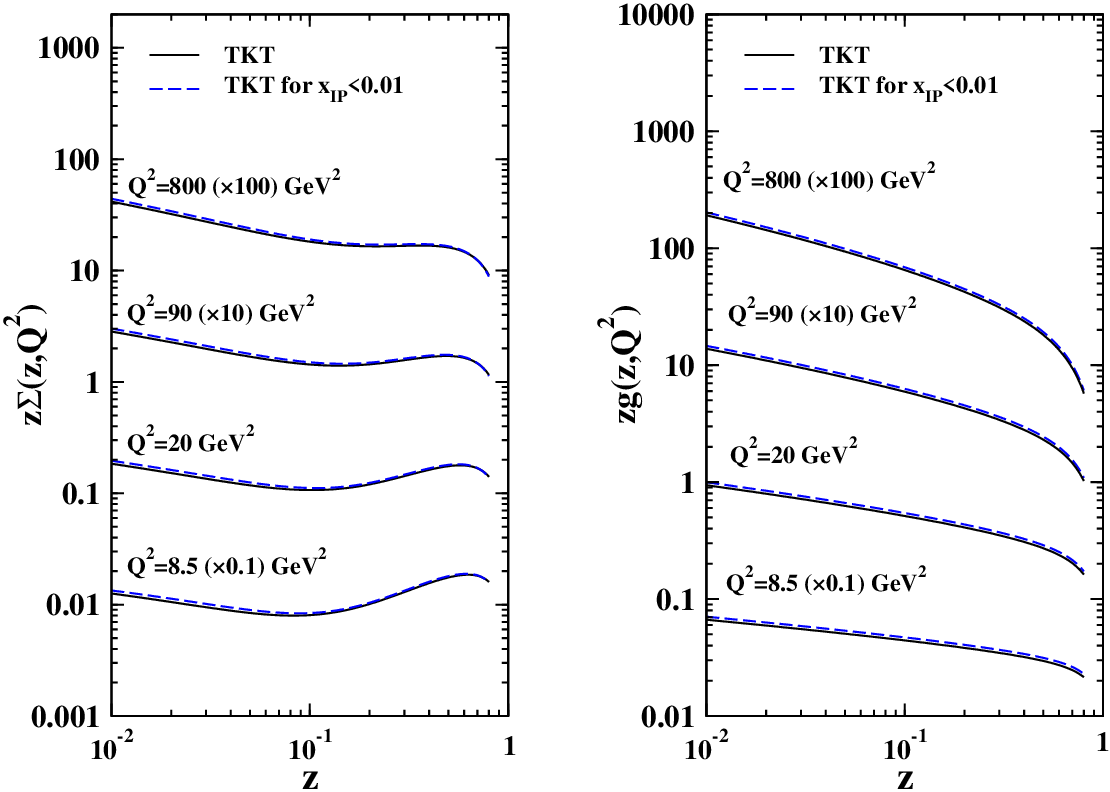}}
\caption{Comparison between the total quark singlet and gluon distributions obtained from our model. Results are presented with no cut on $ \xpom $ (solid) and $ \xpom<0.01 $ (dashed).  }
\label{cuttedxpparton}
\end{figure}

As it was discussed, the contribution of the sub-leading Reggeon trajectories should be considered  for $\xpom$ values substantially larger than $0.01$
. The contribution from these trajectories is modelled using the pion structure function.
The pion PDFs are used in a region of low $\beta$ where they are not directly constrained by data.
There is potential interference between the contributions from the Pomeron and
sub-leading Regge trajectories with vacuum quantum numbers (e.g. the
f-meson, see Eq. (14) of Ref. \cite{Adloff:1997sc}).
In order to limit the influence of sub-leading Reggeon trajectory, we exclude data points with $\xpom>0.01$ and use a single $N_{\reg}$ for all required data sets. 
The results are shown in Figure \ref{cuttedxpparton}. Only very small differences
are observed between both fits. This analysis leads to $\alpha_{\pom}=1.110\pm 0.0041$.

\vspace{1mm} \noindent \\
\begin{table*}[h]
\begin{center}
\begin{tabular}{|c|c|}
\hline \hline Parameters   &  TKT  \\
\hline 
\hline  $A_{\Sigma}$ &$0.17\pm 0.009$  \\
\hline  $B_{\Sigma}$ &$0.08\pm 0.031$   \\
\hline  $C_{\Sigma}$ &$0.53\pm 0.025$  \\
\hline  $D_{\Sigma}$ &$4.88\pm 0.14$  \\
\hline  $E_{\Sigma}$ &$-2.36\pm 0.064$ \\
\hline  $F_{\Sigma}$&$0.30\pm 0.012$  \\
\hline  $A_g$ &$0.44\pm 0.020$   \\
\hline  $\alpha_{\pom}(0)$ &$1.108 \pm 0.0035$  \\  \hline
\hline  $N_{\reg}$(H1-LRG-06) &$(1.22\pm 0.18) \times 10^{-3}$ \\
\hline  $N_{\reg}$(H1-FPS-06) &$(1.21\pm 0.23) \times 10^{-3}$  \\
\hline  $N_{\reg}$(H1-FPS-10)&$(1.36\pm 0.09) \times 10^{-3}$  \\
\hline  $N_{\reg}$(ZEUS-LPS-04) &$(1.64\pm 0.26) \times 10^{-3}$ \\
\hline  $N_{\reg}$(ZEUS-LPS-09)&$(2.25\pm 0.12) \times 10^{-3}$    \\
\hline  $N_{\reg}$(ZEUS-LRG-09)&$(2.19\pm 0.29) \times 10^{-3}$  \\
\hline \hline
$\chi ^{2}/$dof &$601.92/607=0.99$    \\ \hline \hline
\end{tabular}
\end{center}
\caption{{Pomeron quark and gluon densities parameters and their statistical errors for combined data sets in $\overline{\rm MS}$ scheme at the input scale $Q_0^2$=3 GeV$^2$.
\label{tab:parameter}}}
\end{table*}
%
%

\subsection{The method of the minimization and error calculation}
\label{The method of the minimization and error calculation}

The quality of fit is traditionally determined by the $\chi^2$ of the fit to the data \cite{Stump:2001gu}, which is minimized using the MINUIT package \cite{James:1975dr}.
 $\chi_{\mathrm{global}}^{2}$ is defined by
\begin{equation}
\chi_{\mathrm{global}}^{2}(p)=\sum_{i=1}^{n^{data}}\left[\left(\frac{1-{\cal N}_{i}}{\Delta{\cal N}_{i}}\right)^{2}+\sum_{j=1}^{n^{data}}\left(\frac{{\cal N}_{j}\: F_{2,j}^{D,data}-F_{2,j}^{D,theor}(p)}{{\cal N}_{j}\:\Delta F_{2,j}^{D,data}}\right)^{2}\right] \;,\label{eq:chi2}
\end{equation}
where $p$ denotes the set of independent parameters in the fit and $n^{data}$ is the number of data points included. For the $i^{th}$  experiment, $F_{2,j}^{D,data}$, $\Delta F_{2,j}^{D,data}$ and $F_{2,j}^{D,theor}$ denote the data value, measurement uncertainty and theoretical value for $n^{th}$ data point.
$\Delta {\cal N}_{i}$ is the experimental normalization uncertainty and ${\cal N}_{i}$ is an overall normalization factor for the data of experiment $i$. We allow for a relative normalization shift ${\cal N}_{i}$ between different data sets within uncertainties $\Delta {\cal N}_{i}$ quoted by the experiments.

The errors include systematic and statistical uncertainties, being the total experimental error evaluated in quadrature.
We check the fit stability by performing the two approaches with statistical and systematics errors added in quadrature or with statistical errors only. For a given set of data, the results based on the fit with statistical or total errors are very close.
When moving to a combined fit of all data sets, although DPDFs show small differences between both fits, using statistical errors lead to fit with a large $\chi^2$.

There are clear procedures for propagating error experimental uncertainties on the fitted data points through to the PDF uncertainties. The most common is the Hessian approach.  In this case we can consider
\begin{equation} \label{eq:hessian}
  \Delta\chi^2_{\rm global} \equiv \chi^2_{\rm global} - \chi_{\rm min}^2 = \sum_{i,j} H_{ij}(a_i-a_i^{(0)})(a_j-a_j^{(0)}),
\end{equation}
where the Hessian matrix is defined as
\begin{equation}
  H_{ij} = \left.\frac{1}{2}\frac{\partial^2\,\chi^2_{\rm global}}{\partial a_i\partial a_j}\right|_{\rm min}.
\end{equation}
The standard formula for linear error propagation is
\begin{equation} \label{eq:heserror}
  (\Delta F)^2 = \Delta\chi^2 {\sum_{i,j}\frac{\partial F}{\partial a_i}(H_{ij})^{-1}\frac{\partial F}{\partial a_j}}.
\end{equation}
Since the derivative of $F$ with respect to each parameter $a_i$ is required, this formula is not easily calculable.
It can be improved by finding and rescaling the eigen vectors of $H$ \cite{Pumplin:2001ct,Martin:2002aw,Martin:2009iq}. In term of the rescaled eigevectors $z_i$, the increase in $\chi^2$ is given by
\begin{equation} \label{eq:hessiandiag}
  \chi^2_{\rm global} - \chi^2_{\rm min} = \sum_{i} z_i^2.
\end{equation}
The uncertainty on a quantity is then obtained applying
\begin{equation} \label{eq:symmunc}
 ( \Delta F)^2 = \frac{1}{2} \sum_{i} \left[F(S_i^+)-F(S_i^-)\right]^2,
\end{equation}
where $S_i^+$ and $S_i^-$ are PDF sets displaced along eigenvector directions by the given $\Delta \chi^2$.
The uncertainties on our DPDFs following this method are presented in Figure \ref{combined}.\\


%
%
\begin{table}[h]
\begin{center}
\begin{tabular}{|c|c|c|} 
\hline\hline Lable & $\chi^2/$dof & Data points 
\\ \hline\hline
H1-LRG-06 & 0.79 & 190  \cite{Aktas:2006hy}  \\
H1-FPS-06 & 0.53 & 40  \cite{Aktas:2006hx} \\
H1-FPS-10 &0.91 & 100   \cite{:2010kz} \\
ZEUS-LPS-04 & 0.23 &   27  \cite{Chekanov:2005vv} \\
ZEUS-LPS-09 &0.78 & 42  \cite{Chekanov:2008cw}\\
ZEUS-LRG-09 &1.08 & 155  \cite{Chekanov:2008cw}\\
H1-LRG-11 &0.89 &  67  \cite{Collaboration:2011ij} \\
\hline\hline
\end{tabular}
\vspace{2mm} \noindent \\
\caption{{$\chi^2$ values per data set for the gobal QCD fit with statistical and systematic errors added in quadrature. \label{tab:all data}}}
\end{center}
\end{table}

\begin{figure}[t]
\centerline{\includegraphics[width=0.7\textwidth]{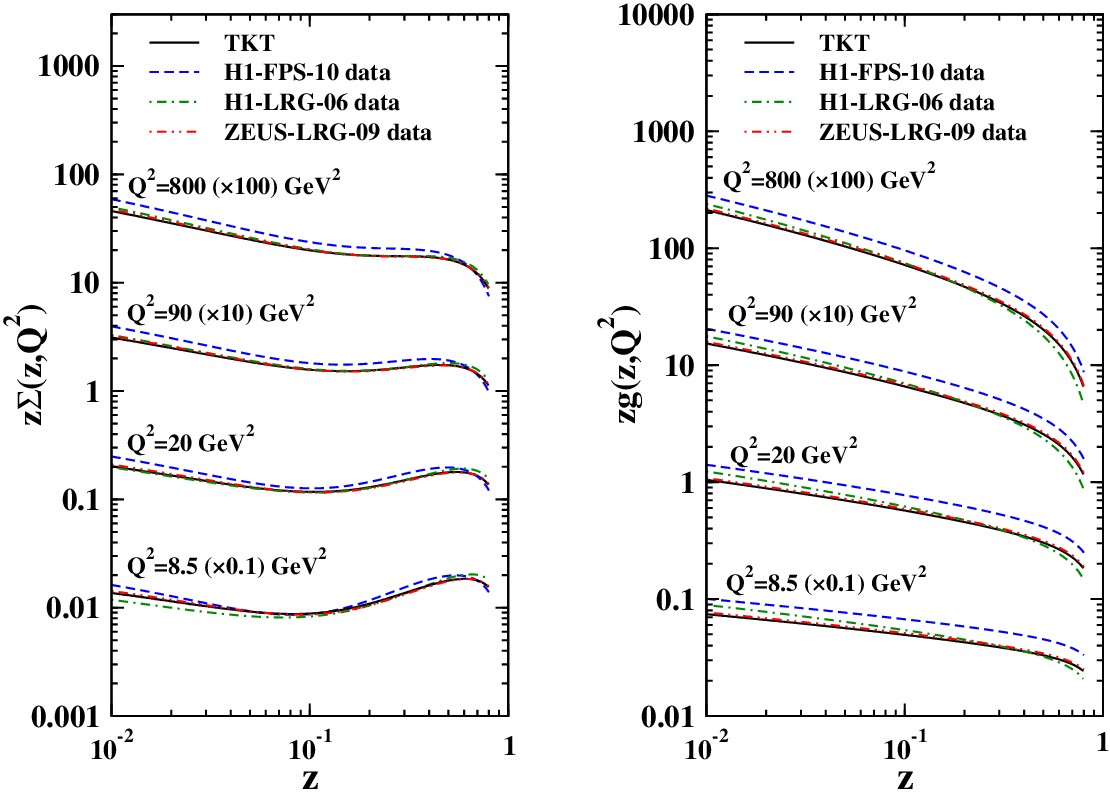}}
\caption{Singlet and gluon distributions of Pomeron as a function of $ z $ derived from QCD fits on H1-FPS-10 \cite{:2010kz} data alone, H1-LRG-06 \cite{Aktas:2006hy} data alone, ZEUS-LRG-09 \cite{Chekanov:2008cw} data alone and all the data sets together.}
\label{allpartons}
\end{figure}

We present $\chi ^{2}$ values for individual data sets in Table~\ref{tab:all data}, which would allow some assessment of the degree of compatibility. It justifies our approach to combine all data sets.
These fit results are  displayed in Figure \ref{allpartons}. We show the quark and gluon densities in the Pomeron for individual H1 and ZEUS data sets.
We note that the Pomeron is gluon dominate for all fits.
\section{ Discussion of fit results}
\label{Discussion of fit results}
\subsection{Behaviour of cross section and  structure function}
\label{Behaviour of cross section and  structure function}
\begin{figure}[b]
\centerline{\includegraphics[width=0.6\textwidth]{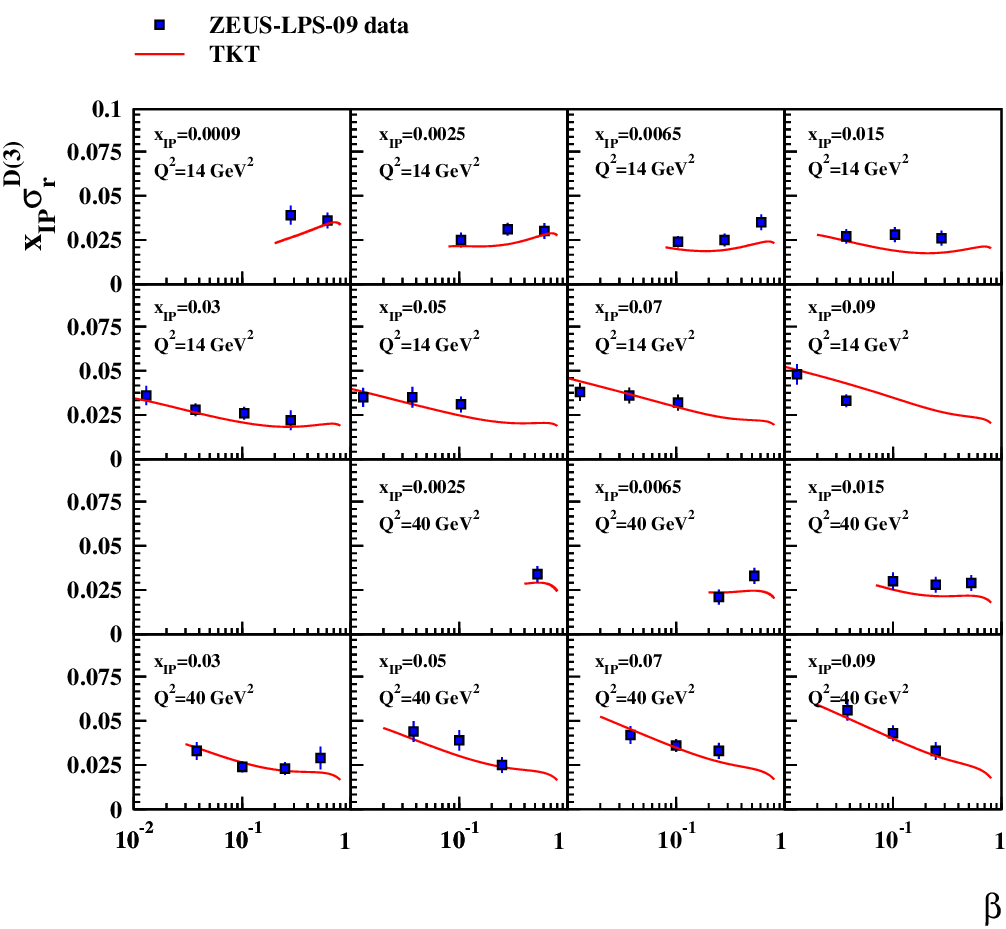}}
\caption[]{The ZEUS-LPS-09  \cite{Chekanov:2008cw} diffractive cross section multiplied by $\xpom$, $\xpom\sigma_r^{D(3)}$, as a function of $\beta$ for different regions of $Q^2$ and $\xpom$. The error bars indicate the statistical and systematic errors added in quadrature. The curves show our model reduced by a global factor1.33  to correct for the contributions of proton dissociation processes as described in Section \ref{Data sets}.}
\label{ZEUSLPS-09}
\end{figure}

\begin{figure}[t] \unitlength 1mm
  \begin{center}
    \begin{picture}(100,161) 
      \put(-25,92){\epsfig{file=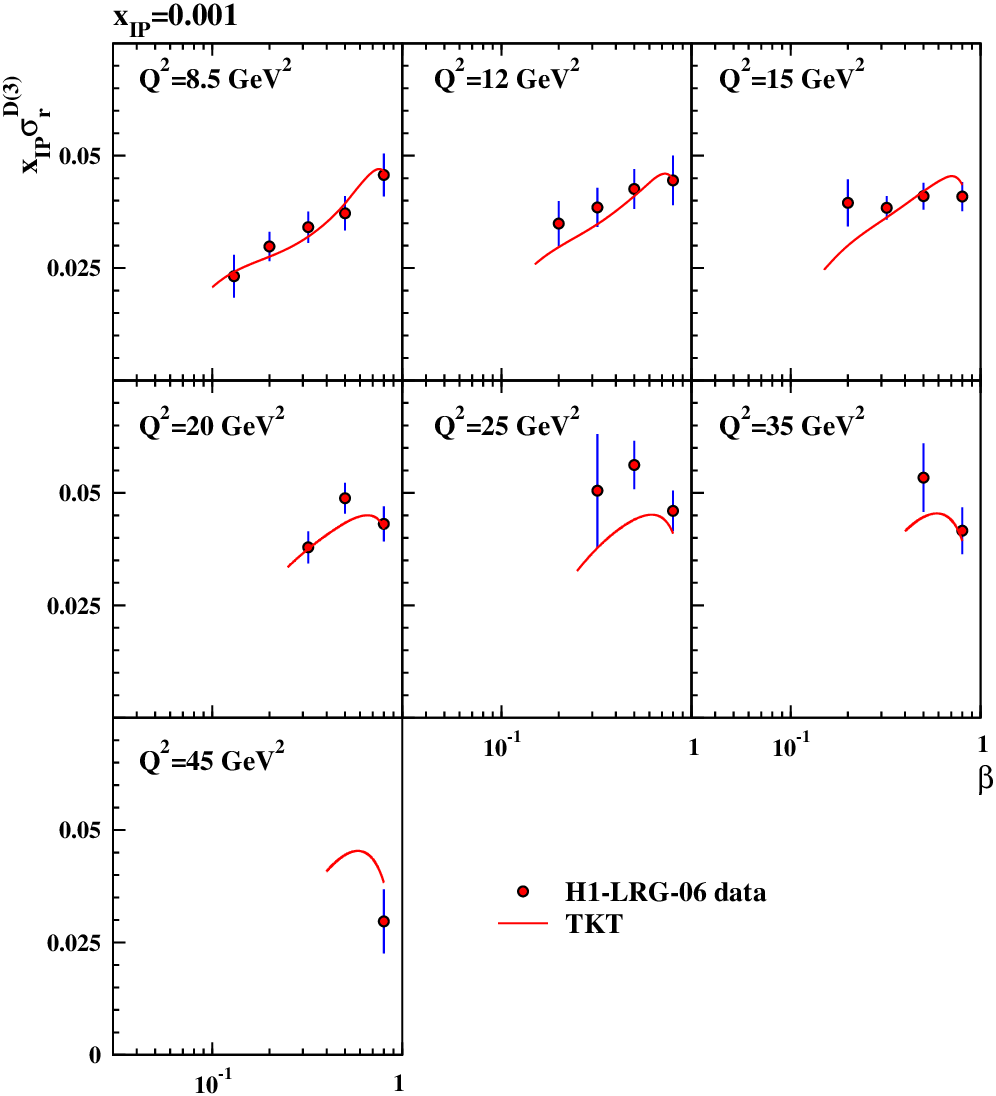,width=0.46\textwidth}}
      \put(55,88){\epsfig{file=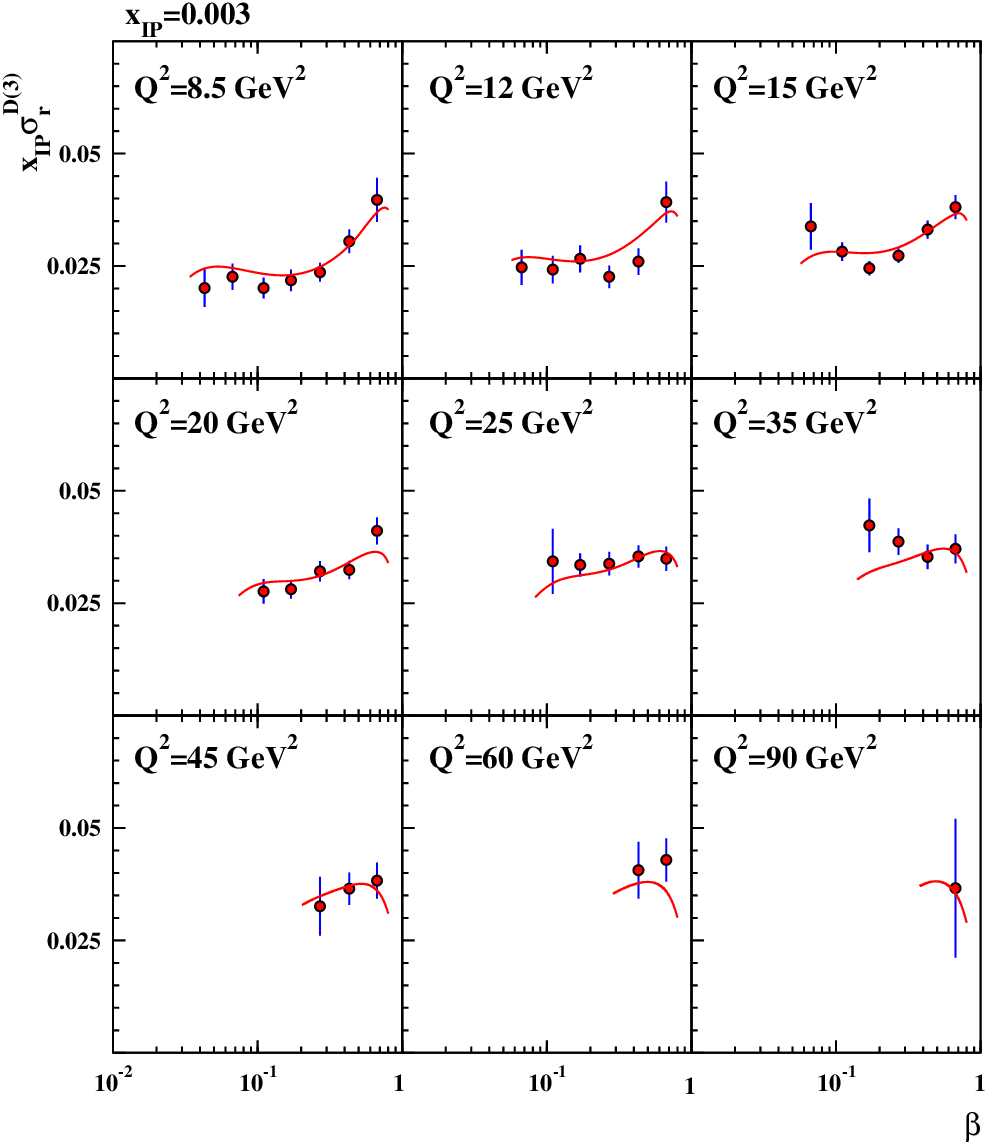,width=0.46\textwidth}}
      \put(-25,-23){\epsfig{file=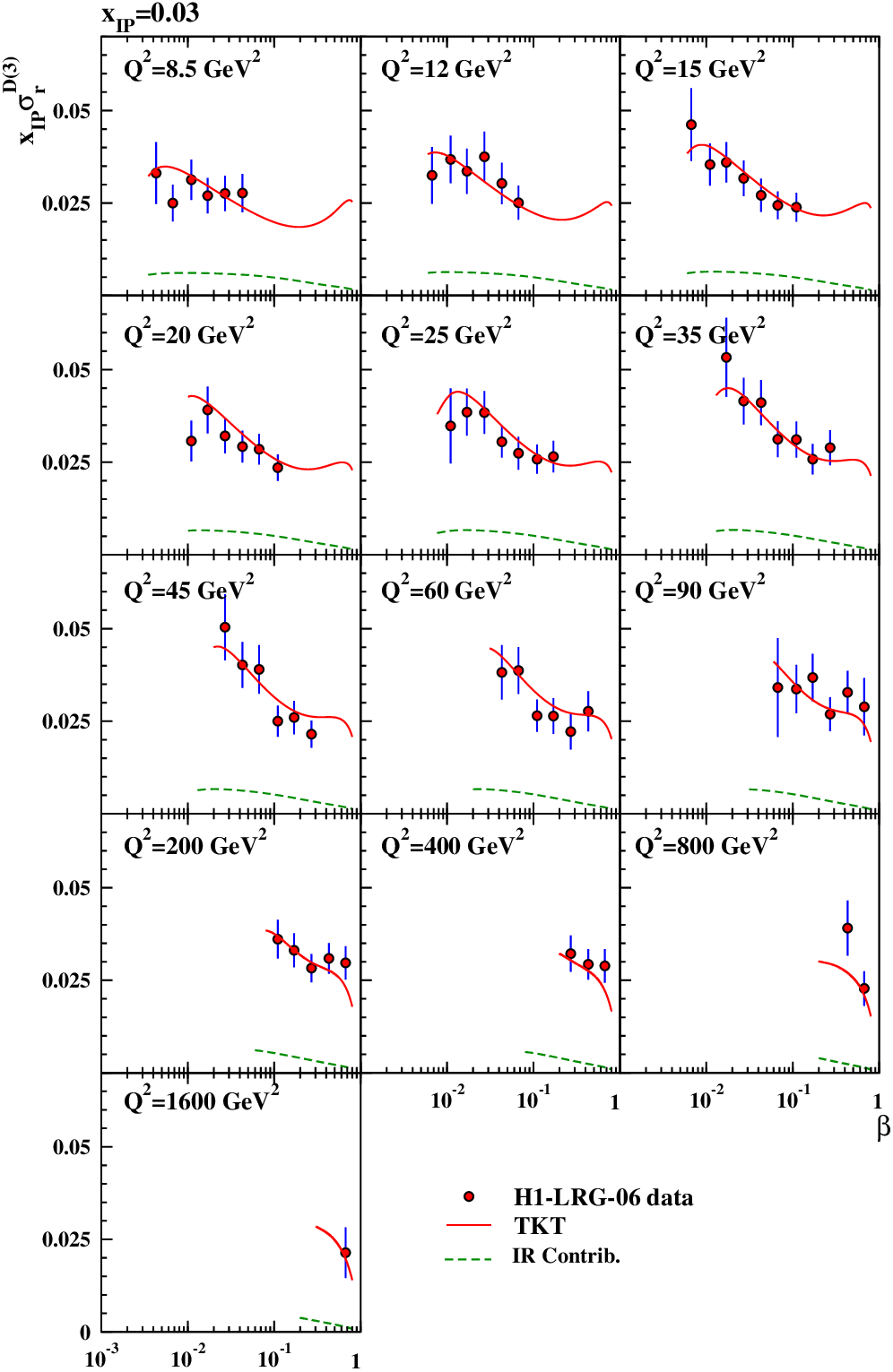,width=0.46\textwidth}}
     \put(55,-23){\epsfig{file=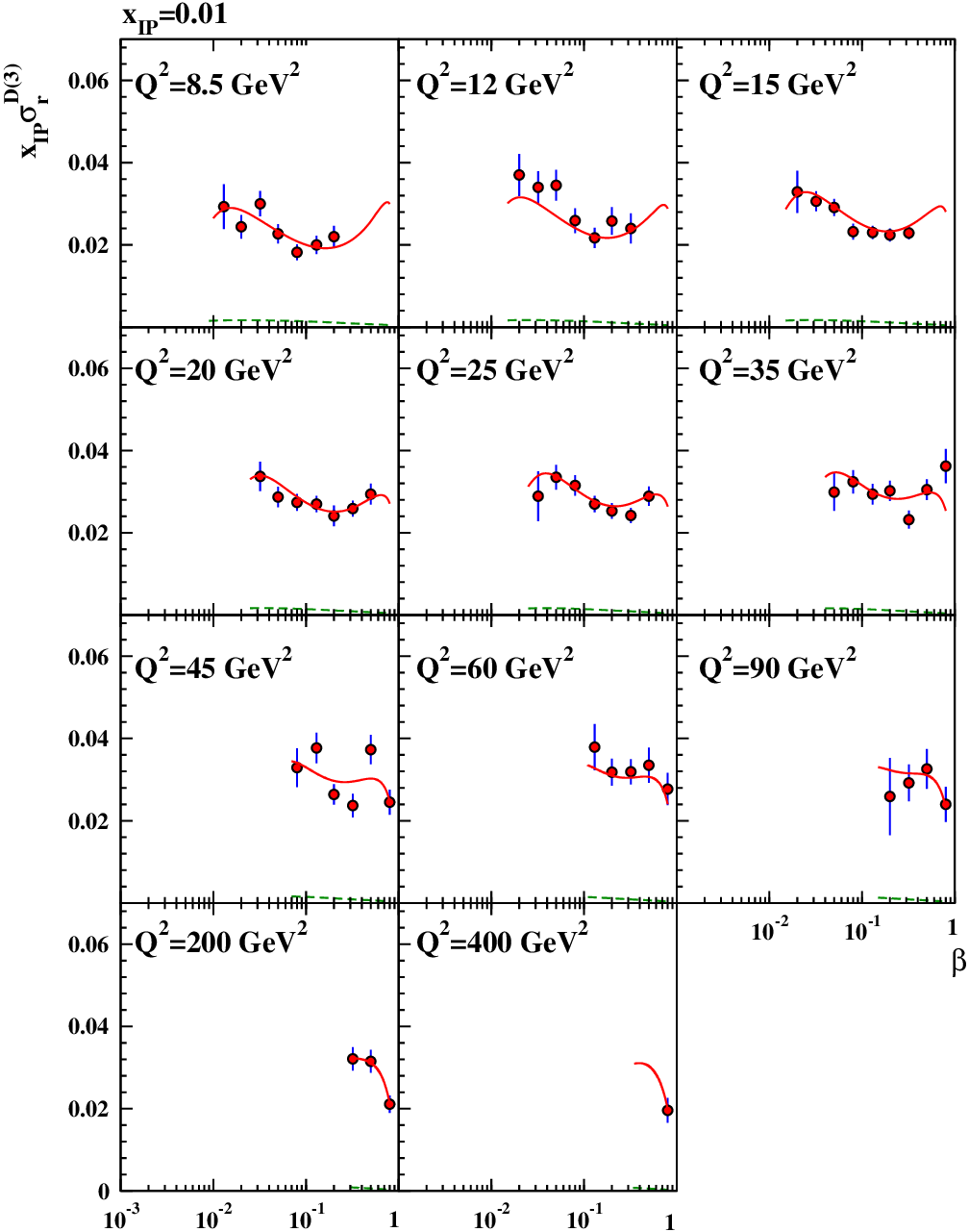,width=0.46\textwidth}}
\end{picture}
  \end{center}
\vspace{8mm} \noindent \\
   \caption[]{The H1-LRG-06 \cite{Aktas:2006hy} reduced diffractive cross section, multiplied by $\xpom$, $\xpom \sigma_r^{D(3)}$, as a function of $\beta$ for different regions of $Q^2$ and $\xpom$. The error bars indicate the statistical and systematic errors added in quadrature. The curves show our model. The contribution of the sub-leading exchange alone is also shown in $\xpom=0.03$ and $0.01$.}
\label{LRG versus beta}
\end{figure}

\begin{figure}[h]
\centerline{\includegraphics[width=0.7\textwidth]{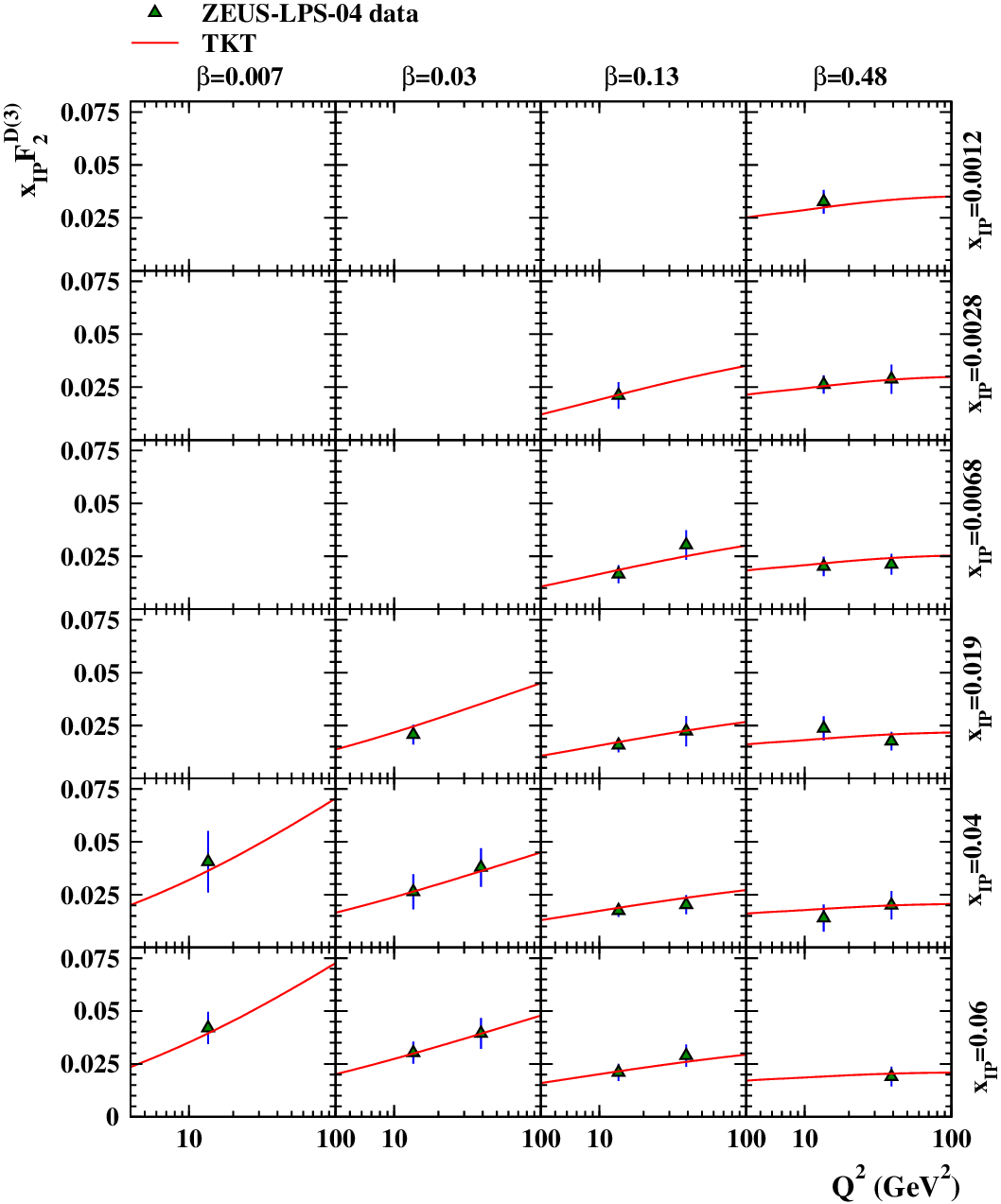}}
\caption[]{The ZEUS-LPS-04 \cite{Chekanov:2005vv} reduced diffractive structure function  multiplied by $\xpom$, $\xpom F_2^{D(3)}$, as a function of $Q^2$ for different regions of $\beta$ and $\xpom$. The error bars indicate the statistical and systematic errors added in quadrature. The curves show our model reduced by a global factor 1.33  to correct for the contributions of proton dissociation processes  as described in Section \ref{Data sets}.}
\label{ZEUS-LPS-04}
\end{figure}

\begin{figure}[t] \unitlength 1mm
  \begin{center}
    \begin{picture}(100,169) 
      \put(-25,82){\epsfig{file=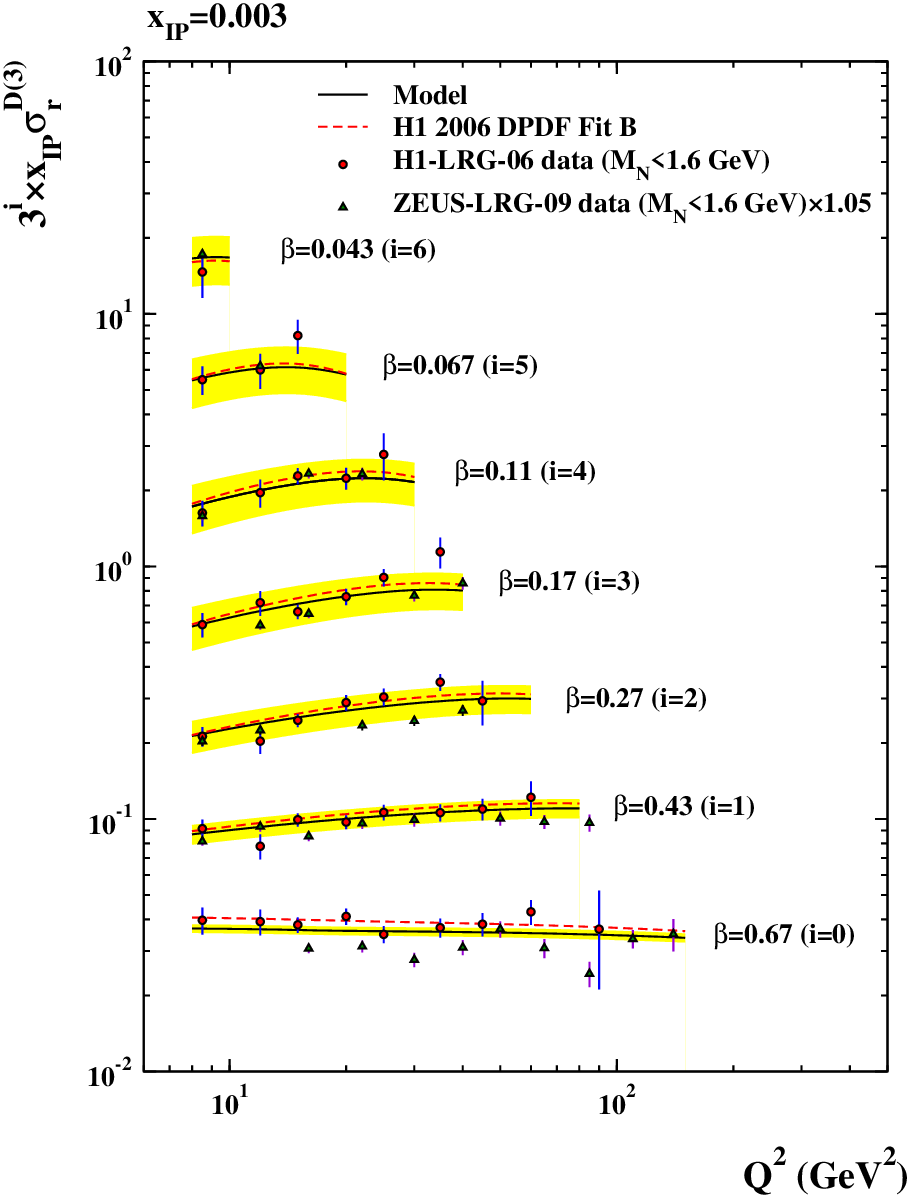,width=0.45\textwidth}}
      \put(55,82){\epsfig{file=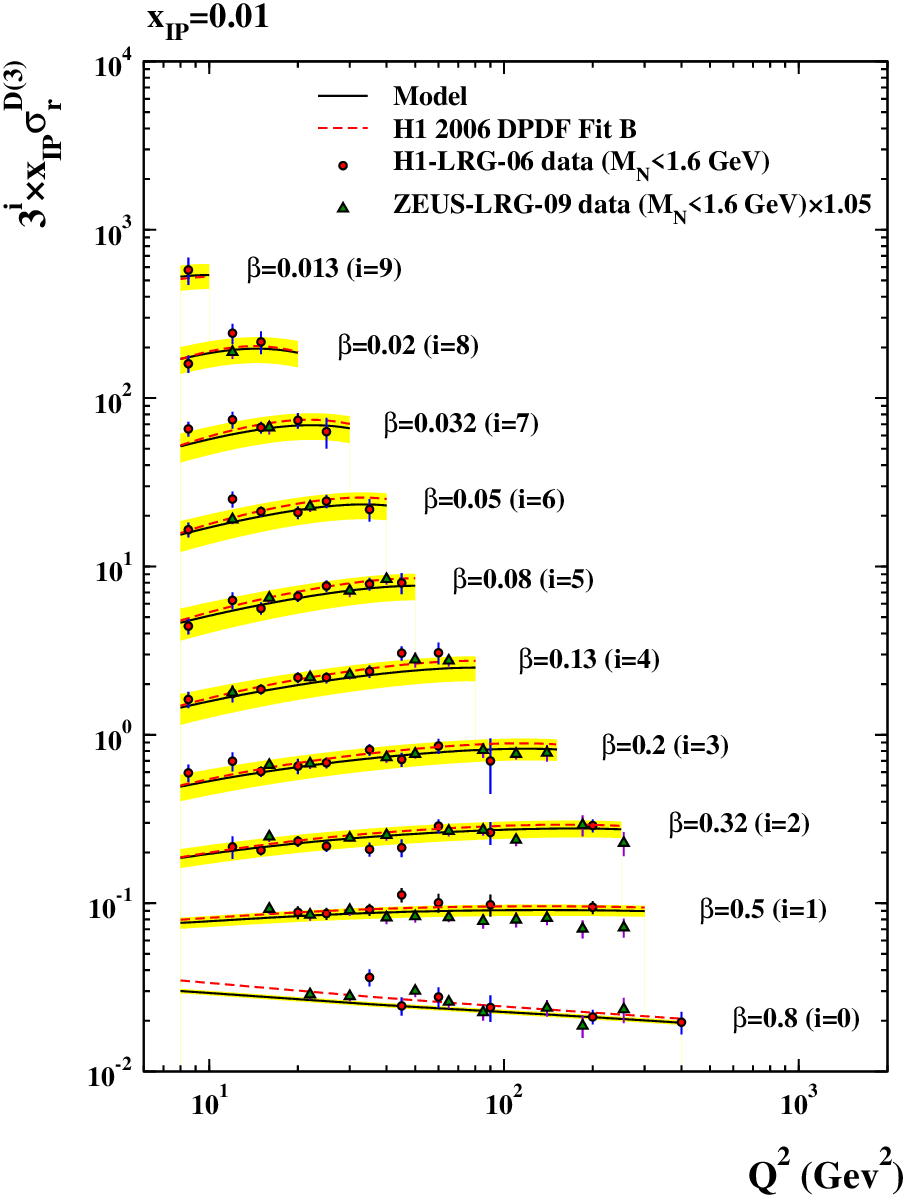,width=0.45\textwidth}}
      \put(-25,-17){\epsfig{file=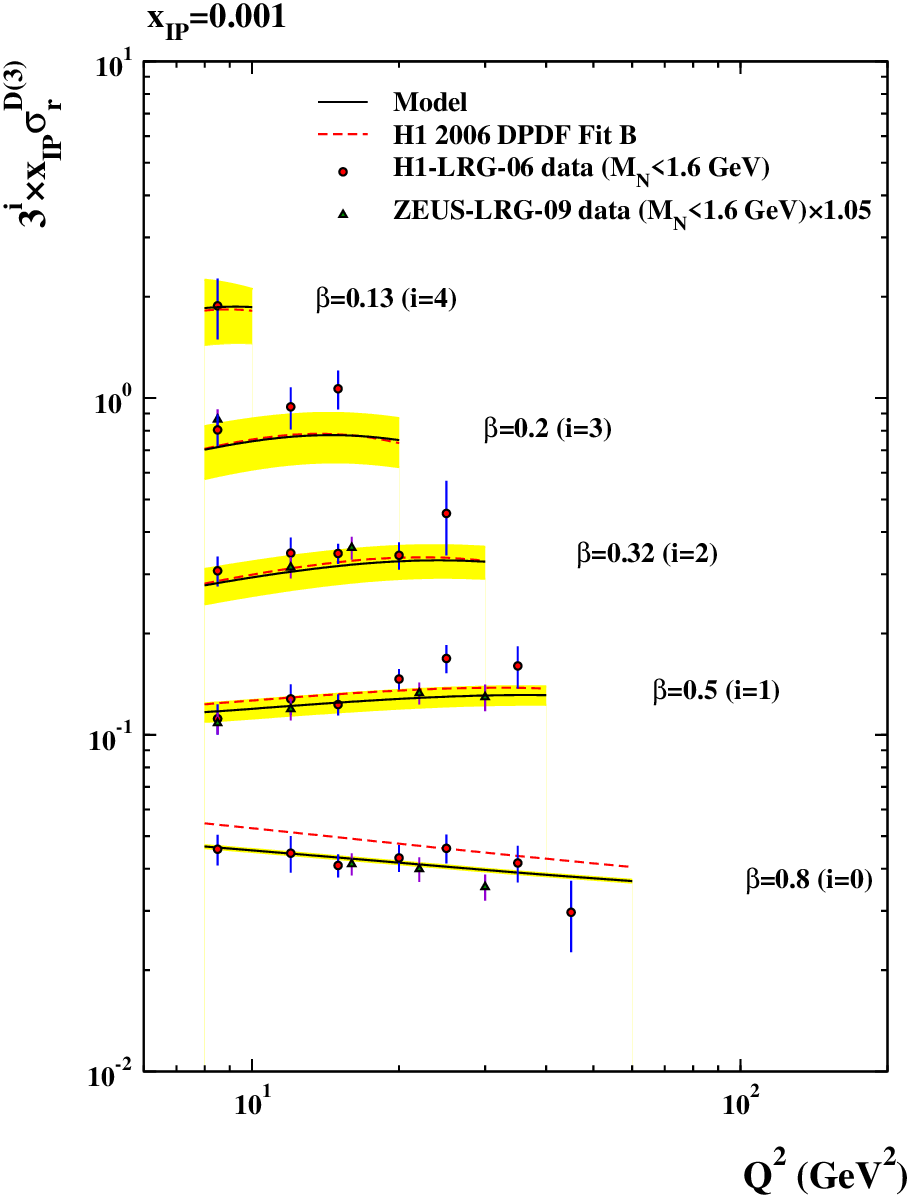,width=0.45\textwidth}}
     \put(55,-17){\epsfig{file=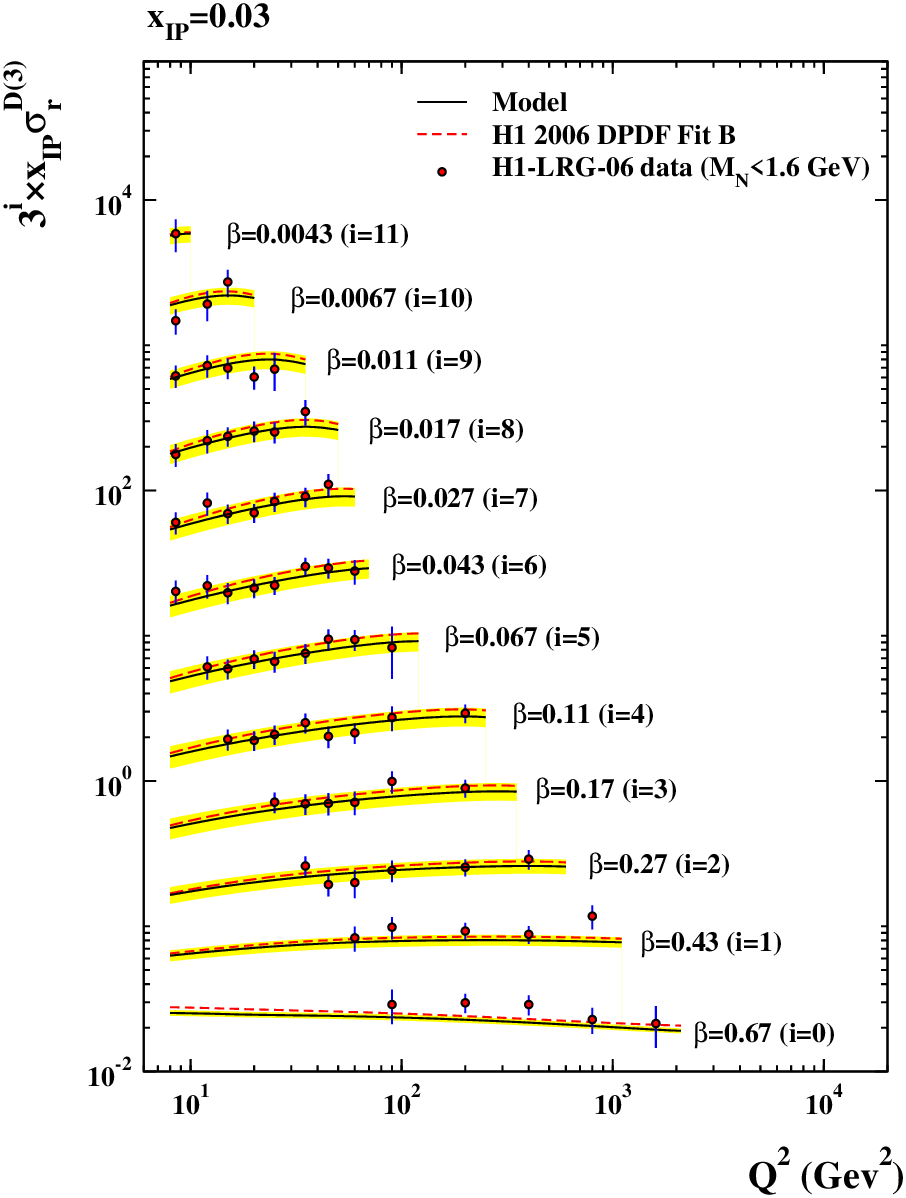,width=0.45\textwidth}}
\end{picture}
  \end{center}
\vspace{0mm} \noindent \\
   \caption[]{Comparison between the H1-LRG-06 \cite{Aktas:2006hy} and ZEUS-LRG-09 \cite{Chekanov:2008cw}  measurements
after correcting the latter data set to $M_Y<1.6 \GeV$  by applying scale factor of 1.05. The measurements are compared with our model (solid curve) and the results
of the H1 2006 DPDF Fit B  (dashed curve) \cite{Aktas:2006hy}, which was based on the H1 data shown.}
\label{fig:comp_lrgg}
\end{figure}

\begin{figure}[h]
\centerline{\includegraphics[width=0.756\textwidth]{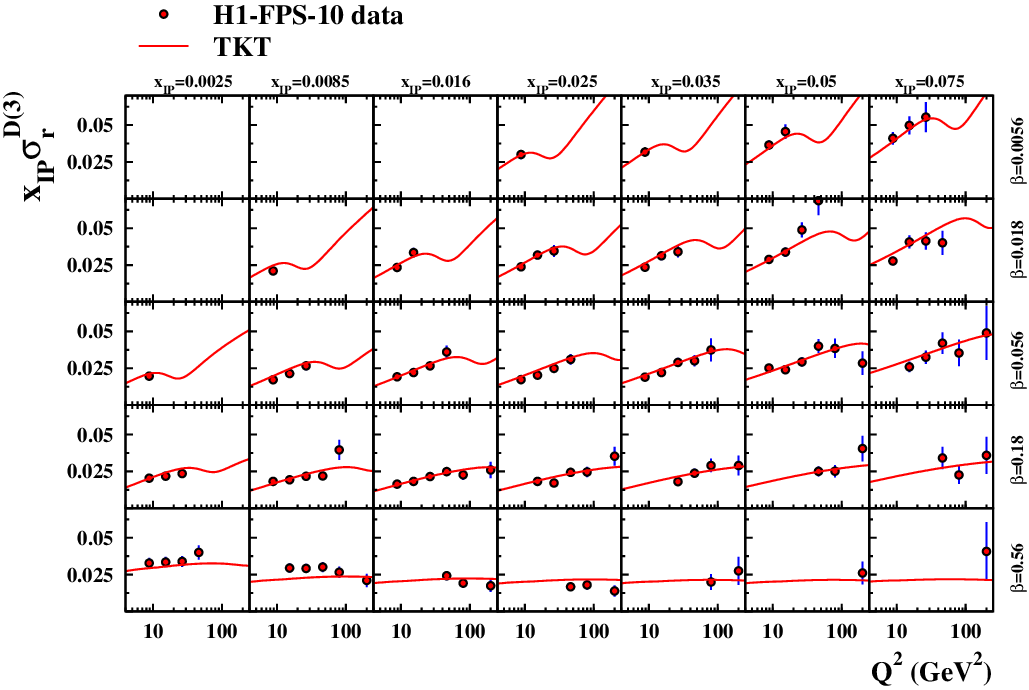}}
\caption[]{The H1-FPS-10 \cite{:2010kz} reduced diffractive cross section, multiplied by $\xpom$, $\xpom \sigma_r^{D(3)}$, as a function of $Q^2$ for different regions of $\beta$ and $\xpom$. The error bars indicate the statistical and systematic errors added in quadrature. The curves show our model reduced by a global factor 1.20 to correct for the contributions of proton dissociation processes  as described in Section \ref{Data sets}.}
\label{H1-FPS-10}
\end{figure}

\begin{figure}[b]
\centerline{\includegraphics[width=0.648\textwidth]{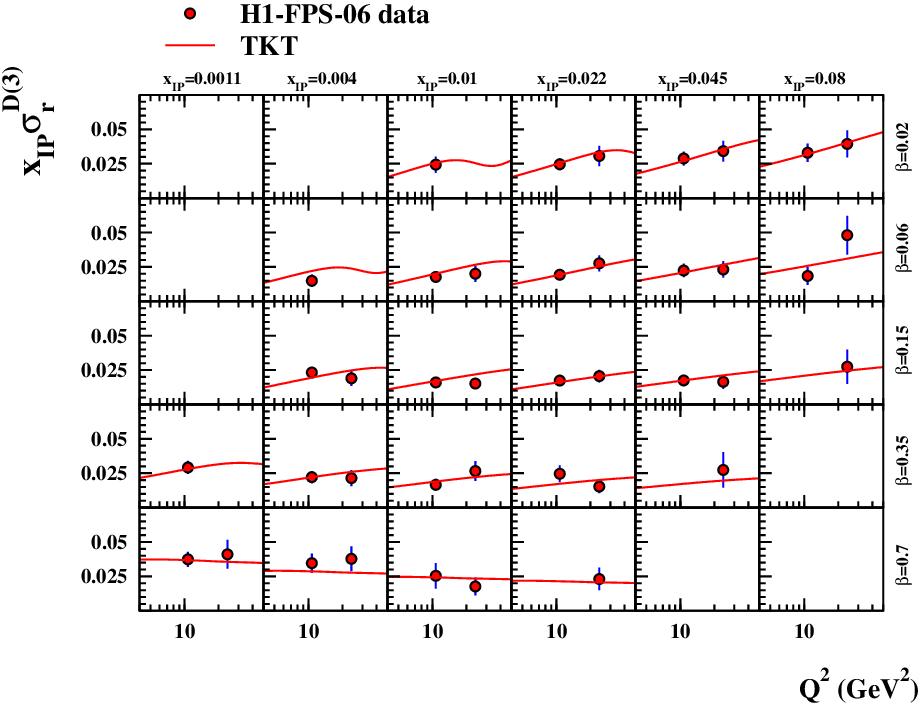}}  %
\caption[]{The H1-FPS-06 \cite{Aktas:2006hx} reduced diffractive cross section, multiplied by $\xpom$, $\xpom \sigma_r^{D(3)}$, as a function of $Q^2$ for different regions of $\beta$ and $\xpom$. The error bars indicate the statistical and systematic errors added in quadrature. The curves show our model reduced by a global factor 1.23 to correct for the contributions of proton dissociation processes  as described in Section \ref{Data sets}.}
\label{H1-FPS-06}
\end{figure}

The data on $\sigma_{r}^{D}$ and $F_{2}^{D}$, which describes the structure of the Pomeron exchanged in the t-channel in diffraction, have two prominent characteristics \cite{Arneodo:2005kd,Alekhin:2005dy}:
\begin{itemize}
\item{Treatment versus $\beta$}

As can be seen in Figures \ref{ZEUSLPS-09} and \ref{LRG versus beta}, diffractive cross section is widely smooth in the measured $\beta$ range. Considering the similarity  between $\beta$ in diffractive DIS and $x_{Bj}$ in inclusive DIS, this is very different from the treatment of $\sigma_r$. Typical cross section effectively reduces for $x_{Bj}\geqslant 0.2$.

\item{Treatment versus $Q^2$}

As illustrated in Figure \ref{ZEUS-LPS-04}, the structure function $F_{2}^{D}$ rises with $Q^2$ for all $\beta$ values (except the highest). This brings the scaling violations of $F_2$ to mind, except that $F_2$ increases with $Q^2$ only for $x_{Bj}\leqslant 0.2$. In the proton, negative scaling violations manifest the existence of the valence quarks, though positive scaling violations are due to the growth of the sea quark and gluon densities. Consequently, the $F_2^D$ data imply that the partons resolved in diffractive  events are mostly gluons. 
 \end{itemize}
These results were already seen in H1 and ZEUS papers in the mid 90's.
Figures \ref{fig:comp_lrgg}, \ref{H1-FPS-10} and \ref{H1-FPS-06} display the reduced diffractive cross section as a function of $Q^2$ for different regions of $\beta$ and $\xpom$.
Our model describes all the data well.

\begin{figure}[h]
\centerline{\includegraphics[width=0.7\textwidth]{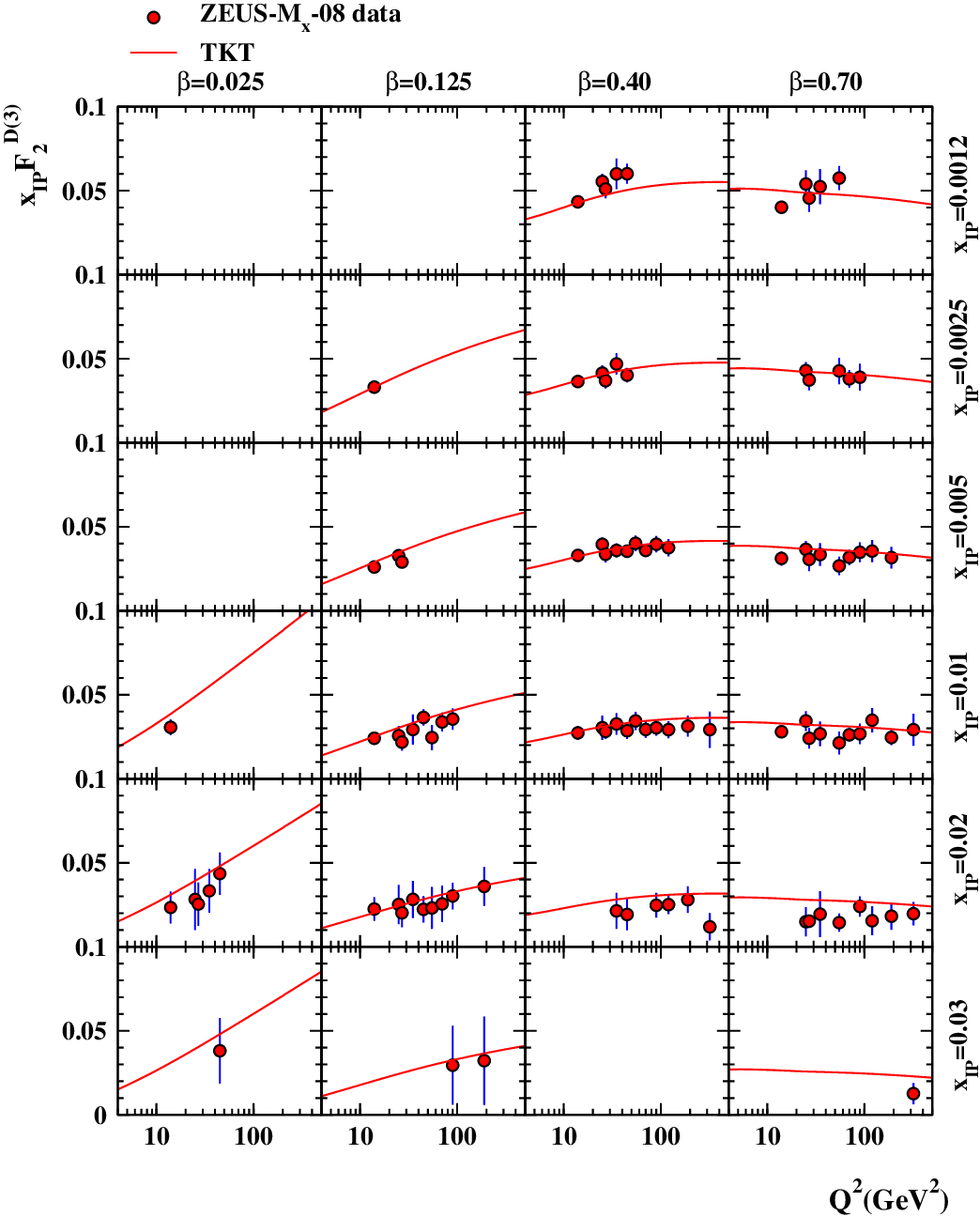}}
  \caption[]{The ZEUS-$M_x$-08 \cite{Chekanov:2008fh} reduced diffractive structure function  multiplied by $\xpom$, $\xpom F_2^{D(3)}$, as a function of $Q^2$ for different regions of $\beta$ and $\xpom$. The error bars indicate the statistical and systematic errors added in quadrature. The curves show our model reduced by a global factor 0.88 to correct for the contributions of proton dissociation processes  as described in Section \ref{Data sets}.}
\label{Zeus-Mx-08}
\end{figure}

\begin{figure}[h]
\centerline{\includegraphics[width=0.7\textwidth]{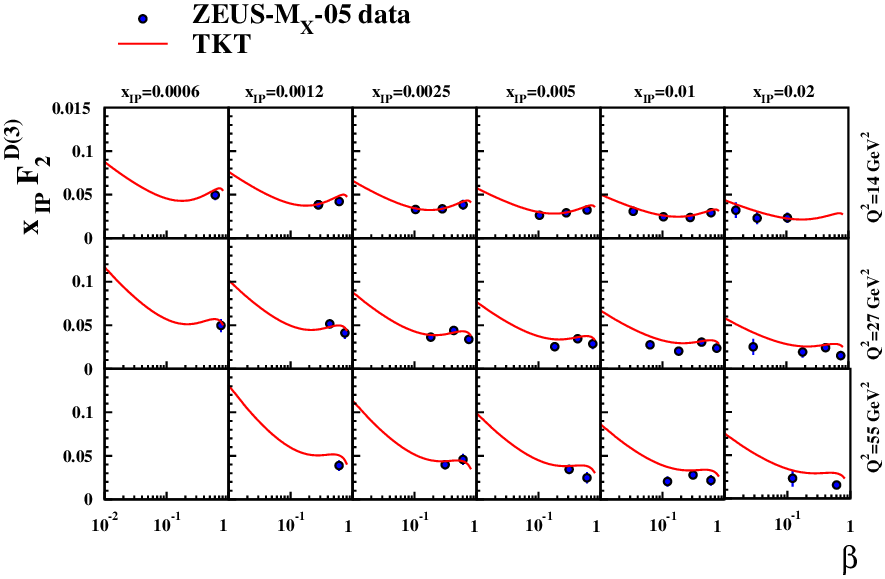}}
\caption[]{The ZEUS-$M_X$-05 \cite{Chekanov:2004hy} diffractive structure function multiplied by $\xpom$, $\xpom F_2^{D(3)}$, as a function of $\beta$ for different regions of $Q^2$ and $\xpom$. The error bars indicate the statistical and systematic errors added in quadrature. The curves show our model reduced by a global factor 0.88  to correct for the contributions of proton dissociation processes as described in Section \ref{Data sets}.}
\label{ZEUS-Mx-05}
\end{figure}

Using the fit results of TKT superimposed on $M_X$ data in Figures \ref{Zeus-Mx-08} and \ref{ZEUS-Mx-05} shows that the fit can describe most kinematic regions. It leads to the conclusion that there seems to be compatibility between all data sets.

\subsection{Longitudinal structure function}
\label{Longitudinal structure function}
Predictions for $F_L^D$ are mainly determined by the form of the gluon density extracted from the fit.
Consequently, the longitudinal structure function provides a way of studying the gluon distribution and a test of perturbative QCD \cite{ICHEP, Newman:2005mm}. This is considerably important at the lowest $z$ values.
\\
Measurements of $F_L^D$ became possible following the reduced proton beam energy runs at the end of HERA operation \cite{H1-prelim} and  first results are recently presented in Ref. \cite{Collaboration:2011ij}.

\begin{figure}[t]
\centerline{\includegraphics[width=0.58\textwidth]{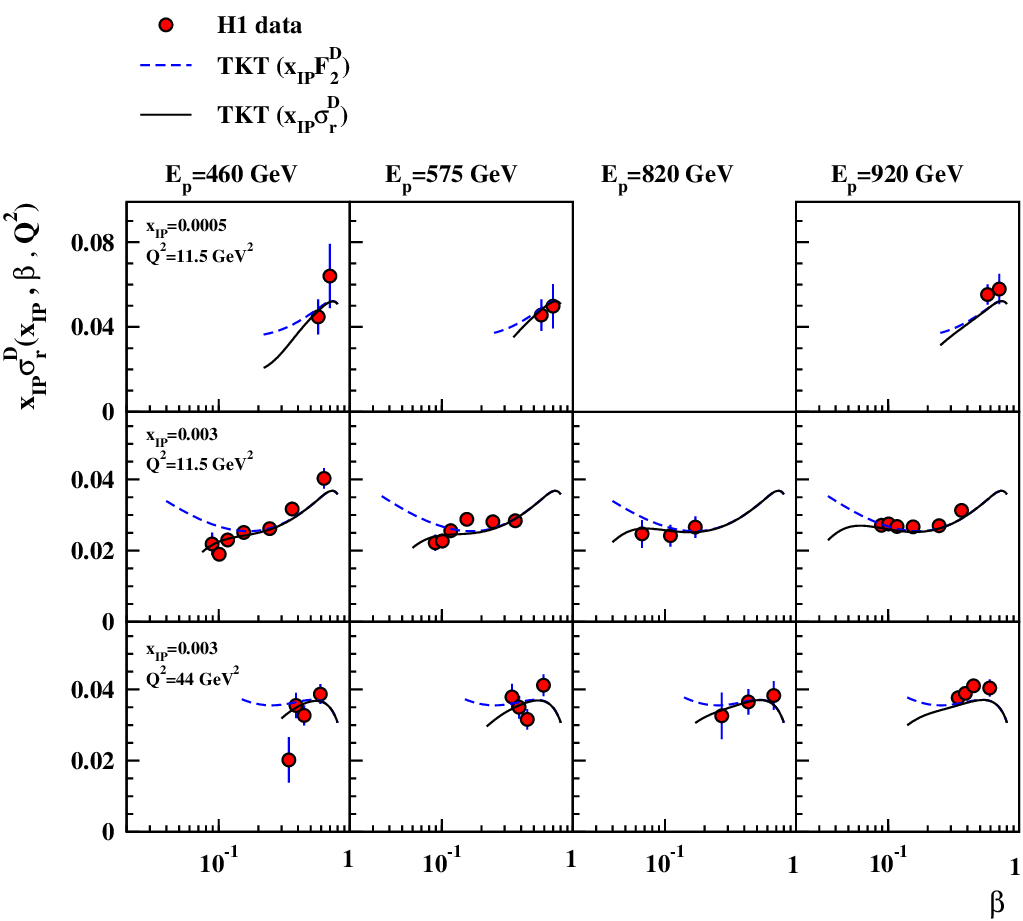}}
\caption[]{The reduced diffractive cross section, multiplied by $\xpom$, $\xpom \sigma_r^{D(3)}$, as a function of $\beta$ at fixed
$Q^2$ and $\xpom$ for (from left to right) the 460, 575, 820 and 920 $\GeV$ data sets \cite{Collaboration:2011ij}. The error bars indicate the statistical and systematic errors added in quadrature. The curves show our model reduced by a global factor 0.97, 0.99 and 0.97 for  $E_p$=460, 575 and 920 GeV, respectively. }
\label{sigmaandF2}
\end{figure}

The reduced diffractive cross section as a function of $\beta$ for the different proton beam energies \cite{Collaboration:2011ij} together with the prediction of our model are shown in Figure \ref{sigmaandF2}.
Discrepancies of the measured cross
sections from our $F_2^D$ curves are considerable in the low and medium energy data.

\begin{figure}[b]
\centerline{\includegraphics[width=0.67\textwidth]{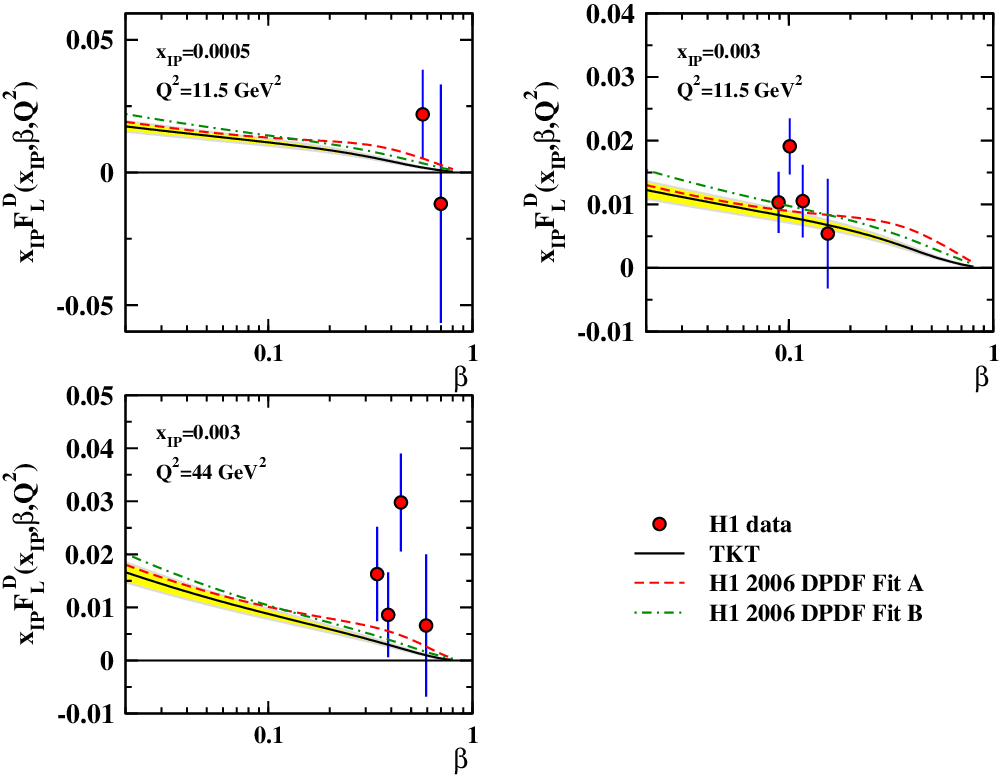}}
\vspace{-0.3cm}
\caption[]{$x_{\pom}F_L^D$ measurement \cite{Collaboration:2011ij} as a function of $\beta$ measured at fixed
$Q^2$ and $\xpom$.
The present fit is the solid curve. Also shown are the results of H1 2006 Fit A (dashed) and and Fit B (dashed-dotted) \cite{Aktas:2006hy}.}
\label{FL}
\end{figure}

The measurements of $F_L^D$ \cite{Collaboration:2011ij} are shown as a function of $\beta$ in Figure \ref{FL}. The data are compared with the predictions of our model and H1 DPDF Fits A and B  \cite{Aktas:2006hy}. All three models are compatible with the data.

\begin{figure}[h]
\centerline{\includegraphics[width=0.6\textwidth]{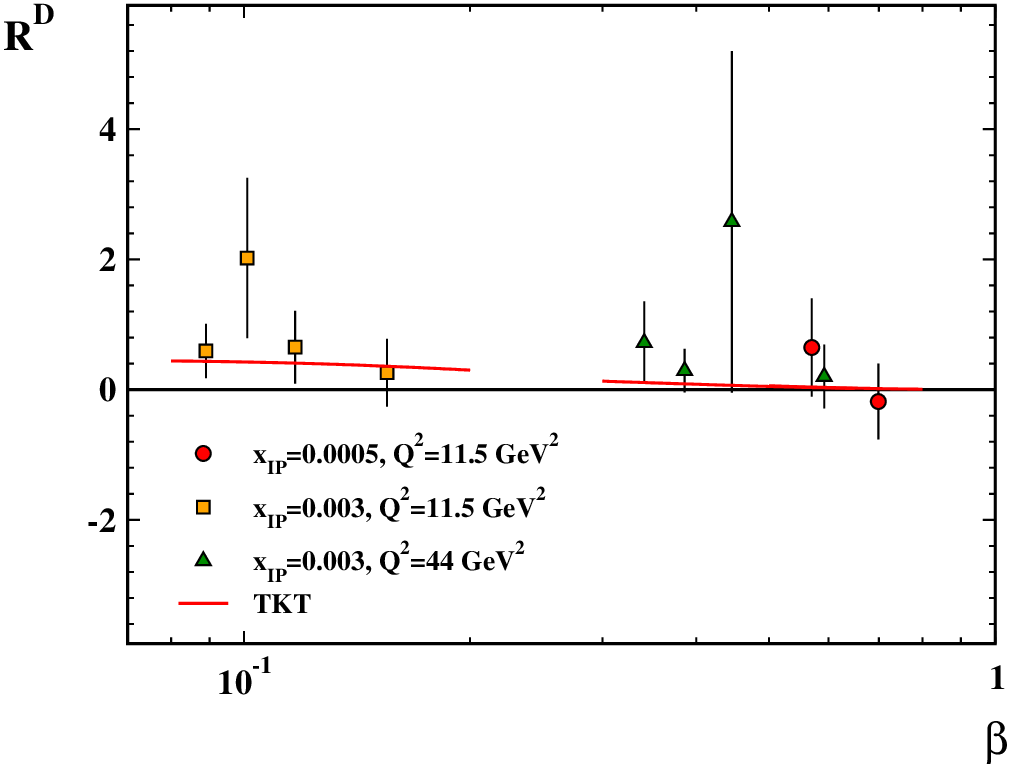}}
\vspace{-0.3cm}
\caption{The ratio measurement \cite{Collaboration:2011ij},  $R^D$, as a function of $\beta$  at the indicated values of $\xpom$ and $Q^2$.
The error bars shown are the statistical and systematic uncertainties added in quadrature. Our result is
the solid curve.}
\label{Ratio}
\end{figure}

The relative sizes of the diffractive cross sections, $R^D=F_L^D/(F_2^D-F_L^D)$, is defined recently \cite{Collaboration:2011ij} by similarity to the inclusive DIS case \cite{:2008tx}.
The measurement of $R^D$ is shown as a function of $\beta$ in Figure \ref{Ratio}. The data are consistent with our model.

\subsection{Charm contribution to the structure function}
\label{Charm contribution to the structure function}
\begin{figure}[b]
\centerline{\includegraphics[width=0.7\textwidth]{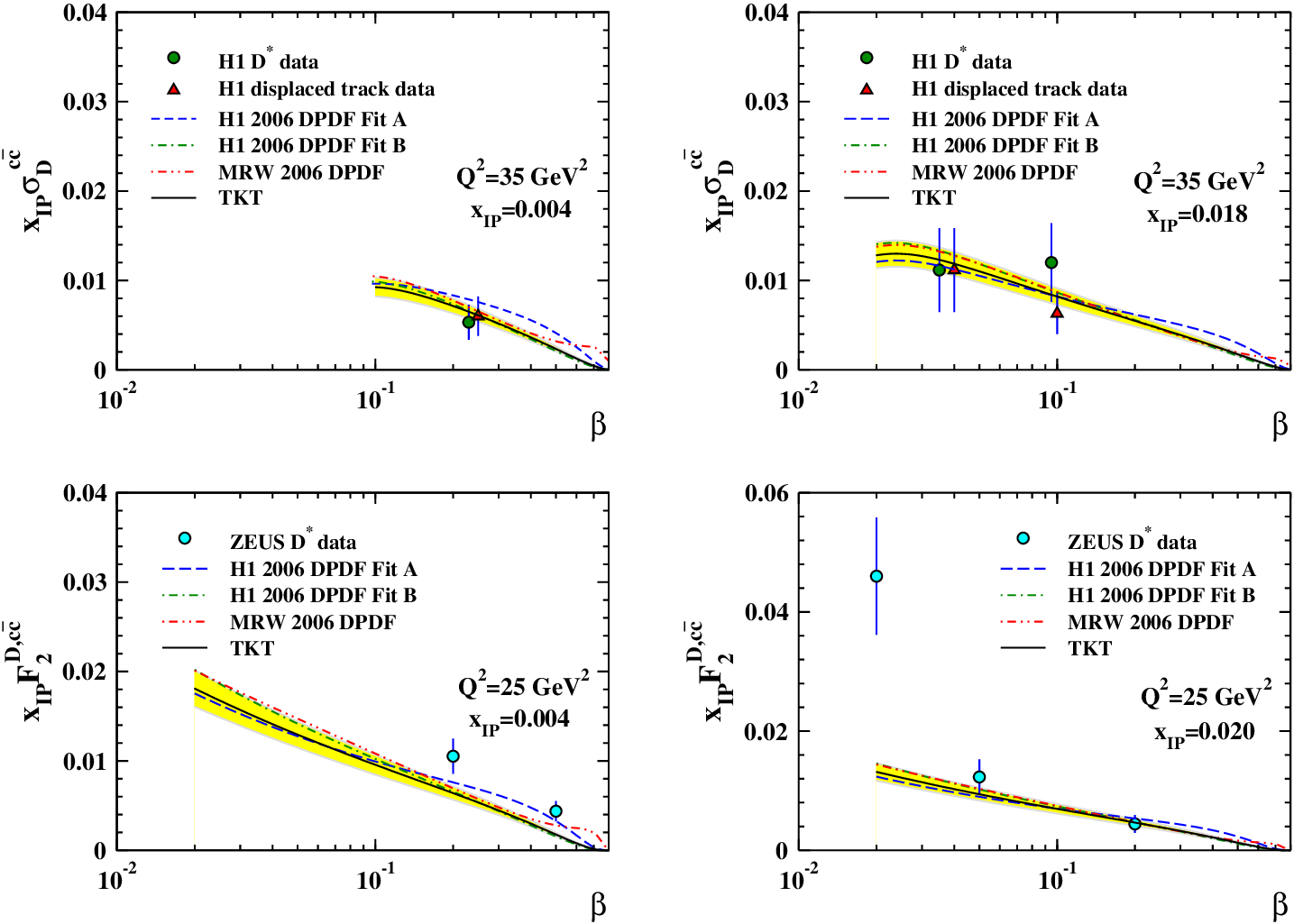}}
\caption[]{Comparison of our result (solid curve) for the contribution of the charm quarks to the diffractive cross section and structure function with H1 2006 DPDF \cite{Aktas:2006hy}  Fit A (dashed curve), Fit B (dashed-dotted curve) and MRW (dashed-dotted-dotted curve) \cite{Martin:2006td} shown as a function of $\beta$ for different values of $\xpom$. The data obtained from the H1 \cite{Aktas:2006up} and ZEUS \cite{Chekanov:2003gt} $D^*$ production and from
 H1  displaced track method \cite{Aktas:2006up}. The error bars of the data points represent the statistical and systematic error in quadrature. Measurements at the same values of $\beta$ are displaced for visibility. }
\label{H1heavy}
\end{figure}

Charm production in diffractive DIS has been measured by the H1 and ZEUS collaborations.
Due to the sensitivity of charm production to gluon-initiated processes this process is very important \cite{McDermott:1996}.
Charm quarks are selected by two independent methods in H1 \cite{Aktas:2006up}. They are selected by the full reconstruction of $D^*$ mesons or by reconstructing the displacement of track, while just the later is used in ZEUS \cite{Chekanov:2003gt}.  In Figure \ref{H1heavy} we present our results for charm diffractive cross section and structure function together with  H1 2006 DPDF \cite{Aktas:2006hy} Fit A, Fit B and  MRW \cite{Martin:2006td} and we compare them with H1 and ZEUS data. ZEUS data and MRW curves are corrected with a factor of 1.23 to account for the difference in the measured range from $M_Y=m_p$ to $M_Y<1.6 \GeV$ \cite{Aktas:2006hx}. The predictions are in fair agreement with the data.

\section{Conclusion }
\label{Conclusion }
We have shown that the diffractive observables measured in the H1 and ZEUS experiments at HERA can be well described by a perturbative QCD analysis which fundamental quark and gluon distributions, evolving according to the NLO DGLAP equations in FFNS, are assigned to the Pomeron and Reggeon exchanges.
This work provides a detailed picture of the Pomeron structure.
In particular, a global analysis of all available data has been performed by introducing a new set of quark distribution form for the Pomeron.
The new functional form developed here contains additional parameters provides flexibility to obtain a proper description of the data. 
 We know that to constrain the gluon, the dijet data is required \cite{:2009qja,Aaron:2010su,Aaron:2011mp} and this work is in progress.
 Having extracted the diffractive PDFs, we compute various diffractive structure functions. In general, we find good agreement with the experimental data, and our results are in accord with other determinations from the literature; collectively, this demonstrates progress of the field toward a detailed description of the Pomeron structure.

Although these data obtained by various methods with very different systematics, they are broadly consistent in the shapes of the distribution throughout most of the phase space.
 This is a very important message from
HERA that DPDFs are well compatible for both experiments.

The new measurements performed by H1 in a more extended kinematic regime 
will allow to further refine the results.
Additionally, the Higgs boson may be produced at the LHC via a diffractive process in which
fast protons are detected. A deeper understanding of diffraction, driven by the
HERA result, could therefore aid in the discovery of the Higgs boson \cite{Capua:2011sm,Boonekamp:2002vg,DeRoeck:2002hk}.

A FORTRAN package containing our diffractive light and heavy structure functions
$F_{2,(\pom,\reg)}^{(light,heavy)}$, $F_{L,(\pom,\reg)}^{(light,heavy)}$ as well as the Pomeron densities
$\Sigma$ and $g$ with their errors at NLO in the $\overline{\rm MS}$ scheme can be found
in \texttt{http://particles.ipm.ir/links/QCD.htm} 
or obtained
via e-mail from the authors. These functions are interpolated
using cubic splines in $Q^{2}$ and a linear interpolation in $\log\,(Q^{2})$. The
package includes an example program to illustrate the use of the routines.

\section*{Acknowledgments}

The authors are especially grateful to A.~ De Roeck and  A.~ Levy for reading the manuscript of this paper, fruitful
suggestions and critical remarks. We thank G.~Watt, M.~ Ruspa, L.~Schoeffel, P.~Newman, F.~Schilling, R.~Polifka and D.~Britzger for many valuable discussions. A.~N.~K. thanks the CERN TH-PH division for its hospitality and support. We acknowledge Semnan University and the School of Particles and Accelerators, Institute
for Research in Fundamental Sciences (IPM) for financially supporting this project.




\newpage
%

\newpage
\newpage
%
\newpage
\vspace{5mm} \noindent \\

\newpage
\vspace{1mm} \noindent \\
\newpage
%
\newpage
\newpage
%
\newpage
\newpage


%


\end{document}